\documentclass[sigconf]{acmart}
\AtBeginDocument{%
  }

\copyrightyear{2025}
\acmYear{2025}
\setcopyright{cc}
\setcctype{by}
\acmConference[ICMI '25]{Proceedings of the 27th International Conference on Multimodal Interaction}{October 13--17, 2025}{Canberra, ACT, Australia}
\acmBooktitle{Proceedings of the 27th International Conference on Multimodal Interaction (ICMI '25), October 13--17, 2025, Canberra, ACT, Australia}
\acmDOI{10.1145/3716553.3750752}
\acmISBN{979-8-4007-1499-3/2025/10}

\usepackage{tabularx}
\usepackage{booktabs}
\usepackage{cleveref}
\usepackage{subcaption}

\newcolumntype{L}[1]{>{\raggedright\let\newline\\\arraybackslash\hspace{0pt}}m{#1}}
\newcolumntype{C}[1]{>{\centering\let\newline\\\arraybackslash\hspace{0pt}}m{#1}}
\newcolumntype{R}[1]{>{\raggedleft\let\newline\\\arraybackslash\hspace{0pt}}m{#1}}
\newcolumntype{Y}{>{\centering\arraybackslash}X}

\begin{document}

\title{Multimodal Quantitative Measures for Multiparty Behaviour Evaluation}

\author{Ojas Shirekar}
\email{o.k.shirekar@tudelft.nl}
\affiliation{%
  \institution{TU Delft}
  \city{Delft}
  \country{The Netherlands}
}

\author{Wim Pouw}
\email{W.Pouw@tilburguniversity.edu}
\affiliation{%
  \institution{Tilburg University}
  \city{Tilburg}
  \country{The Netherlands}
}

\author{Chenxu Hao}
\email{c.hao-1@tudelft.nl}
\affiliation{%
  \institution{TU Delft}
  \city{Delft}
  \country{The Netherlands}
}

\author{Vrushank Phadnis}
\email{vrushank@google.com}
\affiliation{%
  \institution{Google}
  \city{San Francisco}
  \country{USA}}

\author{Thabo Beeler}
\email{tbeeler@google.com}
\affiliation{%
  \institution{Google}
  \city{Zurich}
  \country{Switzerland}}

\author{Chirag Raman}
\email{c.a.raman@tudelft.nl}
\affiliation{%
  \institution{TU Delft}
  \city{Delft}
  \country{The Netherlands}
}

\renewcommand{\shortauthors}{Shirekar et al.}

\begin{abstract}
Digital humans are emerging as autonomous agents in multiparty interactions, yet existing evaluation metrics largely ignore contextual coordination dynamics.  We introduce a unified, intervention-driven framework for objective assessment of multiparty social behaviour in skeletal motion data, spanning three complementary dimensions: (1) synchrony via Cross-Recurrence Quantification Analysis, (2) temporal alignment via Multiscale Empirical Mode Decomposition–based Beat Consistency, and (3) structural similarity via Soft Dynamic Time Warping.  We validate metric sensitivity through three theory-driven perturbations—gesture kinematic dampening, uniform speech–gesture delays, and prosodic pitch-variance reduction—applied to $\approx145$ 30-second thin slices of group interactions from the DnD dataset. Mixed-effects analyses reveal predictable, joint-independent shifts: dampening increases CRQA determinism and reduces beat consistency, delays weaken cross-participant coupling, and pitch flattening elevates F0 Soft-DTW costs. 
A complementary perception study ($N=27$) compares judgments of full-video and skeleton-only renderings to quantify representation effects. Our three measures deliver orthogonal insights into spatial structure, timing alignment, and behavioural variability. Thereby forming a robust toolkit for evaluating and refining socially intelligent agents.
Code available on \href{https://github.com/tapri-lab/gig-interveners}{\color{blue}{GitHub}}.
\end{abstract}



\maketitle
\section{Introduction}~\label{sec:intro}

\begin{figure}[htbp]
    \centering
    \includegraphics[trim={1.15cm, 1.5cm, 2cm, 1cm}, clip, width=\linewidth]{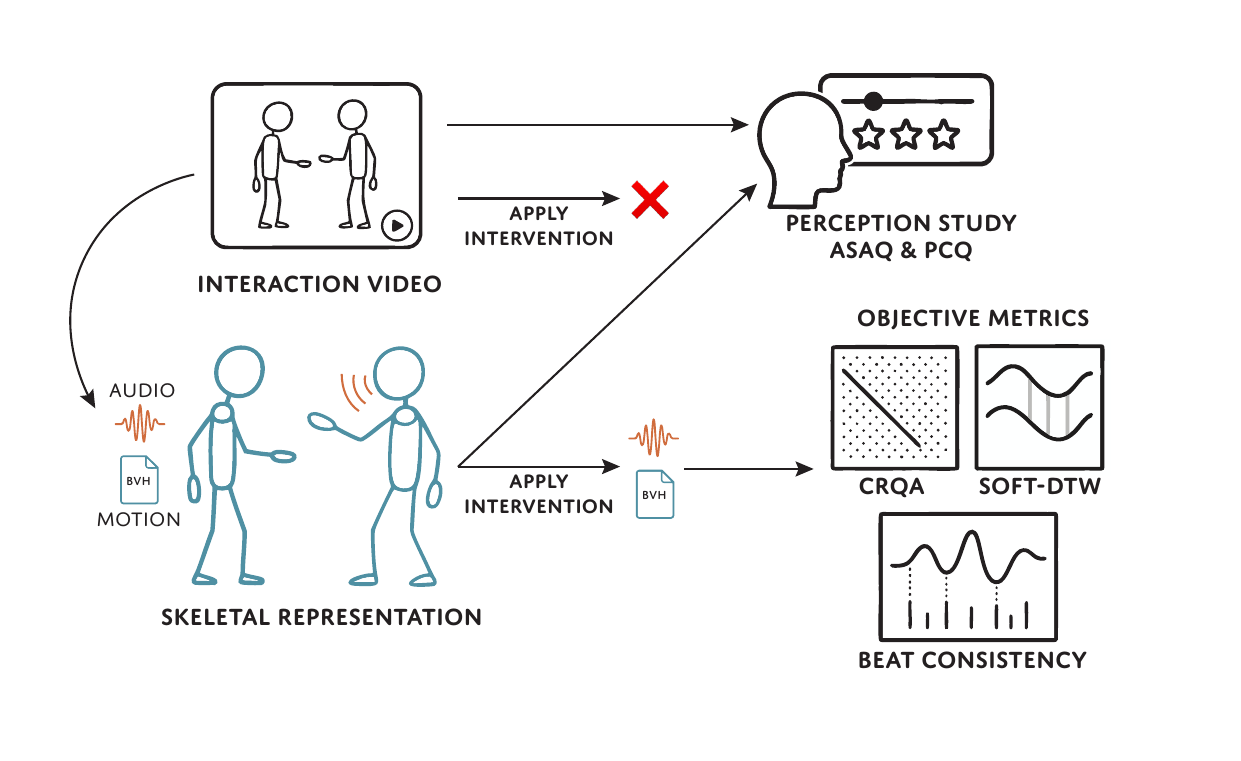}
    \caption{Overview of our evaluation framework. We use skeletal representations of natural interaction videos to enable precise, controlled interventions on kinematics and prosody. These modified embodiments are evaluated using a suite of objective metrics. We also a conduct a user study to gauge the effect of change in representations (video $\rightarrow$ skeleton). Interventions are applied only to skeletons, not to the original videos, allowing causal probing of metric sensitivity to theory-driven perturbations.}\label{fig:teaser-overview}
\end{figure}

Digital humans have already shown promise for applications within the education and mental healthcare domains, such as for evoking curiosity in classrooms~\citep{Paranjape2018-vu}, or eliciting more honest responses in clinical interviews compared to human interviewers~\citep{Lucas2014-kk}. 
In order to develop interactive digital humans to serve as autonomous agents in human-AI collaborative settings, it is crucial to have a consistent and principled framework for evaluating these synthesised behavioural artefacts within interactive social contexts. Crucially, what counts as ``appropriate'' behaviour depends on how it coordinates with the other people in the scene---behaviour is meaningful only in context. In particular, real conversations unfold through fine-grained synchrony: timing and kinematic alignment among participants~\citep{Kendon1970-yw, Kendon1980-qo, Lumsden2014-so, Lumsden2012-zk}. Yet most metrics ignore these behavioural coordination dynamics.

Recent advances in machine learning (ML) render digital humans with unprecedented near-photorealistic fidelity. Yet objective measures that reflect higher-order human perception remain missing~\citep{Nyatsanga2023-pl}, leaving us without a reliable framework for assessing social adeptness.
Therefore, a crucial impediment to further developing ML techniques that enable human-like social dynamics is the absence of a reliable evaluation framework for assessing the quality and social adeptness of generated behaviours in digital humans.

Evaluation of synthesised non-verbal behaviour is inherently challenging. 
One reason is the stochastic nature of behavioural responses: multiple valid responses are possible for the same social stimulus~\citep{Mehrabian1974-hy, Song2023-sp,Raman2021-uz}. 
Moreover, our perception of the validity of those responses is highly subjective~\citep{Nyatsanga2023-pl}.  This challenge is compounded by a mismatch between training objectives and perceptual criteria: the loss functions used to train generative models often optimize for metrics that do not align with how humans judge behavioural quality, meaning that the models may not be learning the very characteristics that drive social believability.

There is still no gold-standard protocol for evaluating synthetic social behaviour.  Subjective studies rely on ad-hoc questionnaires whose items are often collapsed into a single score or lack internal consistency \citep{Wolfert2022-ki}, while ``objective'' work spans everything from mean-squared error to kinematic statistics such as acceleration and jerk \citep{Nyatsanga2023-pl, Kyrlitsias2022-mh}.  The two worlds rarely agree: metrics that look good to a computer frequently fail to match human judgements—an outcome starkly illustrated by the GENEA gesture-generation challenge, where most automatic scores were uncorrelated with perceived human-likeness \citep{Yoon2022-dt}. 
This mismatch reveals a deeper problem, we do not yet know which kinematic or synchrony patterns give rise to higher-order impressions like naturalness or rapport.  Lacking that mapping, the community has no principled way to choose either the right metric or the diagnostic perturbations needed to test it in perception studies.  The situation is further complicated by the fact that nearly all existing metrics treat each actor in isolation and thus ignore the multiparty coordination that defines real conversation.

Taken together, these gaps point to the need for a unified framework that (i) respects the social context in which behaviour unfolds, (ii) is sensitive to multiparty coordination, and (iii) is validated through targeted interventions.  
Most existing metrics were designed for single-actor motion or for speech-gesture appropriateness in isolation~\citep{Yoon2022-dt}; they rarely ask whether the cues are coherent with the surrounding partners’ actions or with the agent’s social goals~\citep{Nyatsanga2023-pl}.  
Even seemingly simple examples underscore the limitation: a jump is meaningless in isolation, expressive during a dance, and disruptive in the middle of a group conversation\textemdash illustrating that social relevance is highly context-dependent~\citep{Grice1975-cr}.  
Recent attempts to incorporate context (e.g.\ \citep{Song2023-sp} for facial reactions) are promising but remain modality-specific and do not generalise to whole-body, multiparty interaction.


Most state-of-the-art behaviour-generation systems operate on \textit{skeletal} motion data rather than generating full video. Joint rotations or 3-D keypoints serve as the primary representation for training losses, diffusion steps, or transformer tokens. Consequently, evaluation must ``speak the same language'' as the models—otherwise, we risk optimising for signals that the evaluation metric cannot access. The use of skeletal representations brings two key advantages. First, it removes appearance cues (faces, clothing, background) allowing us to isolate the kinematic features that underpin interpersonal coordination. This makes controlled manipulations—such as damping hand velocity or delaying gesture onset—both tractable and visually coherent. Second, by presenting observers with the same skeleton-only embodiment that models generate, we can directly assess how well purely kinematic information maps to higher-order perceptual measures such as human-likeness or conversation quality. This leads to our central question: \emph{Can objective metrics applied to skeletal motion  capture the theory-driven kinematic perturbations we introduce?} Addressing this question helps clarify the limits of skeleton-based evaluation and informs the design of metrics that are both aligned with model representations and perceptually meaningful.
In this work, we contribute the following: 
\begin{enumerate}
    \item \textbf{Design} three controlled interventions that systematically perturb gesture kinematics, speech–gesture timing, and vocal pitch.
    \item \textbf{Introduce} a compact panel of cross-modal metrics for multiparty skeleton data.
    \item \textbf{Conduct} an exploratory perception study comparing video recordings to skeletal embodiments to understand how representation modality affects judgments of conversation quality and human likeness in group settings
    \item \textbf{Analyse} the sensitivity of each metric to our interventions in natural group interaction.
\end{enumerate}

\section{Related Work}\label{sec:related-work}

\subsection{Generating Multiparty Social Behaviour}\label{ssec:rel-work-generating}

Recent advances have turned to deep generative approaches for social behaviour synthesis. The synthesis of co-speech gesture generation, for instance, is notoriously challenging due to the highly idiosyncratic, non-repetitive nature of human gestures and their diverse communicative functions~\cite{Nyatsanga2023-pl}. Nevertheless, the field has seen a surge of interest thanks to larger multimodal datasets and powerful models~\citep{Nyatsanga2023-pl}. Modern methods can take various input modalities - e.g. speech audio, text transcripts, or other signals - to drive gesture production~\citep{Nyatsanga2023-pl}. For speech-to-motion generation, early deep-learning models ranged from recurrent networks and GANs to normalizing flows~\citep{Kucherenko2020-if,Ahuja2020-mg}. More recently, diffusion models have gained prominence as social diffusion models for motion. Diffusion-based frameworks address the trade-off between motion quality and diversity that plagued earlier deterministic approaches. For example,~\citet{Dabral2022-ox} introduced a denoising diffusion model to generate co-speech gestures, and \citet{Mughal2024-pi} extended this to a multi-modal conversational diffusion system that produces realistic listener and speaker gestures in coordination. These methods can produce a distribution of plausible motion sequences rather than a single deterministic gesture, better reflecting the variability of human behaviour.

Researchers have also begun exploring language-conditioned or speech-conditioned generation of full-body social behaviour.
Speech-driven approaches such as~\citet{Yoon2020-uf} learn an end-to-end mapping from raw speech features (mel-spectrograms) to upper-body pose sequences for humanoid robots, optionally fusing speech text as an auxiliary cue. ZeroEGGS~\citep{Ghorbani2023-kk} remains speech-audio conditioned, but adds zero-shot style control that can be keyed with a short example motion clip, yielding diverse gestures that synchronize with the spoken signal.
In contrast, text-to-motion frameworks such as MotionCLIP~\citep{Tevet2022-vi} map natural-language prompts into a CLIP~\citep{Radford2021-cs} aligned latent space to drive 3-D body motion. While the Human Motion Diffusion Model~\citep{Tevet2022-vs} can generate long, varied motions from textual action labels or free text. Most of these methods still target a single performer; the recent work of~\citet{Sun2024-rd} moves a step further by jointly generating holistic 3-D motions for both a speaker and a listener in dyadic conversations, conditioning on audio and text and enforcing mutual influence between the two characters. In another line, social behaviour forecasting has emerged to predict how group interactions will unfold.~\citet{Raman2021-uz} emphasize that individuals adapt their behaviour differently across groups depending on contextual factors like relationships and rapport.
This stochastic forecasting of multiparty non-verbal cues (e.g. who will gesture or look at whom next) represents a forward-generative view of social behaviour. This progress sets the stage for evaluating how \textit{socially appropriate} and human-like these generated behaviours truly are.

\subsection{Evaluation Techniques}\label{ssec:rel-work-eval-tech}


Assessing generated social behaviour is challenging because it must account for both physical realism and social meaning. Prior work combines objective metrics such as joint position error, velocity differences, and Fréchet Distance—with subjective human judgments. Although kinematic measures capture low-level fidelity, they often miss context: a gesture may closely match a reference yet still feel awkward or inappropriate with respect to other conversation partners. As a result, user studies remain the gold standard, where participants rate naturalness, appropriateness, or human-likeness of behaviours~\citep{Neff2010-mi,Kucherenko2021-zm}, and instruments like the Godspeed questionnaire gauge perceived personality and emotional impact in human–robot interactions~\citep{Bartneck2023-nq,Deshmukh2018-ut}. For example,~\citet{Neff2010-mi} demonstrated that adding or altering gestures can change perceptions of an agent’s personality, but such studies are time-consuming and hard to reproduce consistently.

A clear gap between objective and subjective metrics has been identified.~\citet{Wolfert2022-ki} reviewed evaluation practices for gesture generation in embodied agents and noted a lack of reliable automatic metrics for social quality. Often, objective scores like mean error or diversity have weak correlation with human judgments of appropriateness or engagement. This gap has motivated work on better proxies for social naturalness. One approach is to measure interpersonal coordination signals. For example, interpersonal synchrony is a desirable emergent quality in multi-party behaviour. Researchers have quantified synchrony via movement correlation and recurrence analyses – e.g.~\citet{Shockley2003-po} showed people’s body sway becomes coupled when they are interacting smoothly. Recent studies go further to link such measures with social outcomes:~\citet{Ruhlemann2024-oh} found that more expressive hand gestures can induce stronger physiological synchrony (skin conductance coupling) between storytellers and listeners, suggesting an objective avenue to gauge a gesture’s social efficacy. Similarly, metrics from dynamical systems like cross-recurrence quantification~\citep{Wallot2016-gj,Wallot2018-kf,Wallot2019-be} have been applied to evaluate the temporal coordination between agents’ multimodal signals.

Context-dependent appropriateness recognizes that, rather than a single ``correct'' response, an AI may produce multiple valid behaviours for the same social cue~\citep{Raman2021-uz}. The Multiple Appropriate Facial Reaction Generation challenge~\citep{Song2023-sp} highlights this by allowing diverse listener expressions and evaluating them with context-sensitive metrics such as the rank correlation of generated versus real facial features (FRCorr) and an expression-space distance (FRDist). While these measures move beyond pure visual realism, they still compare only distributions of synthetic reactions. Ultimately, the true test is whether people perceive multiparty behaviours as natural and believable. Recent studies have begun validating that higher synchrony scores align with greater observer-rated rapport, or that an appropriateness metric predicts user preference between gesture variants~\citep{Miles2009-gm}. The field is rapidly moving beyond low level fidelity measures toward social evaluation criteria such as diversity, synchrony and context awareness, while grounding these quantitative assessments in human judgments through frameworks that combine signal analysis with subjective evaluation of behaviour~\citep{Nyatsanga2023-pl}. 

\section{Methodology}\label{sec:methodology}
We selected CRQA, multiscale beat consistency, and Soft-DTW because together they give a comprehensive, interpretable window into the rich, dynamic structure of social behaviour that simpler statistics miss. CRQA lets us quantify both linear and non-linear synchrony—including transient entrainment and leader–follower dynamics—by mapping when two participants’ state-space trajectories return to similar regions. The multiscale beat consistency score homes in on the critical cross-modal timing between gesture and speech at multiple temporal scales, capturing how co-speech gestures tune prosodic perception and narrative flow. And Soft-DTW provides a flexible, differentiable distance metric that aligns elastic sequences—whether 3D gesture paths or F0 contours—so we can compare natural timing variations within and across individuals.

Crucially, these measures complement one another: CRQA pinpoints when and how long participants are coupled, beat consistency captures precise cross-modal timing, and Soft-DTW quantifies shape similarity under elastic alignment. Integrating them lets us triangulate synchrony across phase, scale, and form for a robust, multi-perspective characterization of social coordination.

\subsection{Behavioural Measures}\label{ssec:behavioural-measures} 

\paragraph{CRQA}

Human social interactions unfold as complex, time-varying sequences of movement, in which partners continuously adapt to one another’s subtle kinematic cues. Traditional linear measures—like cross-correlation or coherence—capture only stationary, time-lagged similarities and often miss the rich, non-stationary coupling that underlies real‐time synchrony. Cross-recurrence quantification analysis (CRQA) overcomes these limitations by reconstructing each participant’s state-space and then directly mapping when and for how long their trajectories return to similar regions. This makes CRQA uniquely sensitive to both linear and non‐linear coordination patterns, robust to noise and differences in signal scale, and capable of capturing transient episodes of entrainment that standard methods overlook~\citep{Wallot2016-gj,Wallot2018-kf,Wallot2019-be}.  Moreover, by examining shifts of the recurrence structures off the main diagonal, CRQA naturally handles phase shifts and time lags in gestures—quantifying not only whether two signals synchronize but also when one leads or follows the other.
 
CRQA has been used for analysing social communication and behavioural characteristics in general and in multi-person group settings~\citep{Wallot2018-kf, Coco2020-xy, Shockley2003-po, Orsucci2013-ny, Tomashin2022-wj}. In our work, we are using CRQA to help quantify synchrony within virtual social interactions. Specifically we shall be making use of the recurrence rate (RR), \% determinism (\%DET) and mean length of recurrence (MeanLR). Typically in a recurrence analysis, the RR is fixed to a reasonable degree through adjusting the recurrence radius (e.g. \(2\%\)) ~\citep{Wallot2019-be}. The recurrence radius defines the maximum distance allowed between two points for them to be considered recurrent. While RR is adjusted to have a certain variance, since it is fixed for the entire dataset it is a valuable output measure of CRQA; indicating the presence recurrences (suggesting presence of attractors in state-space). \%DET  quantifies the proportion of recurrence points forming the diagonal of the recurrence matrix, a higher \%DET value implies a higher structure within the two signals and therefore making it more predictable.

\paragraph{Multiscale Beat Consistency Between Speech and Gestures} 
It is quite well known that human beat gestures are important co-speech factors that tune the perception of prosodic information, narrative structure, and affecting a much wider range of perceived social dynamics~\citep{Pouw2020-qn, Pouw2022-mj, Ter_Bekke2024-uy,Vila-Gimenez2019-zj, Wagner2014-gi}. Previous studies show that humans perceive differences in the temporal relationships between beat gestures and speech, while also affecting their speech perception and memory~\citep{Leonard2011-hw,Bosker2021-sr,Treffner2008-am,Nirme2024-kj}. There is further biomechanical research showing that beat gestures directly couple to the acoustics of the voice through respiration~\citep{Pouw2022-mj, Pouw2020-qn}. 
Given this important entanglement of gesture and speech on multiple functional levels it is likely an important perceptual correlate of human(-like) social behaviour. 
To investigate this multi-scale gesture-speech coordination we turn to empirical mode decomposition~\citep{Rehman2010-td,Rilling2003-ch}. 
Empirical Mode Decomposition (EMD) is an adaptive time-frequency data analysis method that decomposes a signal into a set of Intrinsic Mode Functions (IMFs) representing different frequency components (i.e., different temporal scales), which have been used to characterize speech~\citep{Weston2024-vw,Tilsen2013-yc}. Unlike Fourier analysis, EMD makes no prior assumptions about the data, enabling effective analysis of non-linear and non-stationary signals by separating oscillatory modes dynamically as they change over time. Non-stationary frequency compositions are naturally present in social behaviour obviating predefined basis functions. EMD is therefore a good fit for studying multi-modal signals, which in our case concerns an overall total angular speed of hand-gesture-relevant joints (elbow and wrist joints for left and right), and the smoothed amplitude envelope for speech (following~\citep{Tilsen2013-yc}).
To measure the consistency of timing between the different (EMD-derived) multimodal signals we utilize the beat consistency score as provided in~\citep{Li2021-zf}, whereby the temporal alignment between signal onsets (i.e., beats) is calculated by a guassian-weighted proximity score ($1 =$ perfect synchrony) that is normalized by the number of beats. 
Note that next to application for between-modality within-person analysis, we also apply our multiscale beat consistency metric for cross-person analysis to assess social coordination (e.g., synchrony) patterns (which is known to be multiscale in nature too; e.g.,~\citep{Bigand2024-ux,Abney2014-ml}).

\paragraph{Soft-DTW (SDTW)}
Both the 3D trajectories of a gesture and the F0 contour of speech exhibit rich temporal variability: one person may linger on a hand-wave while another rushes through it; a speaker may stretch or truncate a rising intonation depending on turn-taking cues~\citep{Wichmann2012-lm}. Soft-DTW lets us robustly match these sequences, aligning corresponding sub-gestures or pitch accents even when their durations differ. By focusing on the shape of the motion or pitch contour, rather than rigid clock time, it captures both intra-person consistency (e.g. before/after an intervention) and inter-person coordination (e.g. speaker–listener synchrony). Robust to brief tracking or pitch-tracking artifacts, Soft-DTW gives us a stable, interpretable distance measure that reflects how closely two behaviours (spatio-temporal signals) follow the same dynamic pattern under natural timing shifts.

Soft-DTW extends classic Dynamic Time Warping by replacing the hard minimum over warping paths with a ``soft‐min'' controlled by a smoothing parameter $\gamma$~\citep{Cuturi2017-js}. For $\gamma >0$, the loss is differentiable everywhere, making it compatible with gradient‐based learning and end-to-end modelling of social signals. As $\gamma$ grows, the alignment aggregates more paths and becomes increasingly smooth (down-weighting local misalignments); as $\gamma \rightarrow 0$, it converges to the exact DTW cost. A lower Soft-DTW cost thus indicates tighter temporal similarity after accounting for elastic warping. In practice, we exploit this metric both to compare 3D gesture trajectories, quantifying how an intervention preserves or alters natural motion. We also use it to measure how an induced pitch-variance intervention departs from each speaker’s original F0 contour. Beyond pairwise comparisons, Soft-DTW supports clustering, anomaly detection, and the computation of barycenters, enabling population-level analyses of multimodal social behaviours.

\subsection{Intervention Friendly Motion Representation}\label{ssec:data-pipeline}

\begin{figure}[t]
     \centering
     \begin{subfigure}[t]{0.48\linewidth}
         \centering
         \includegraphics[width=\textwidth]{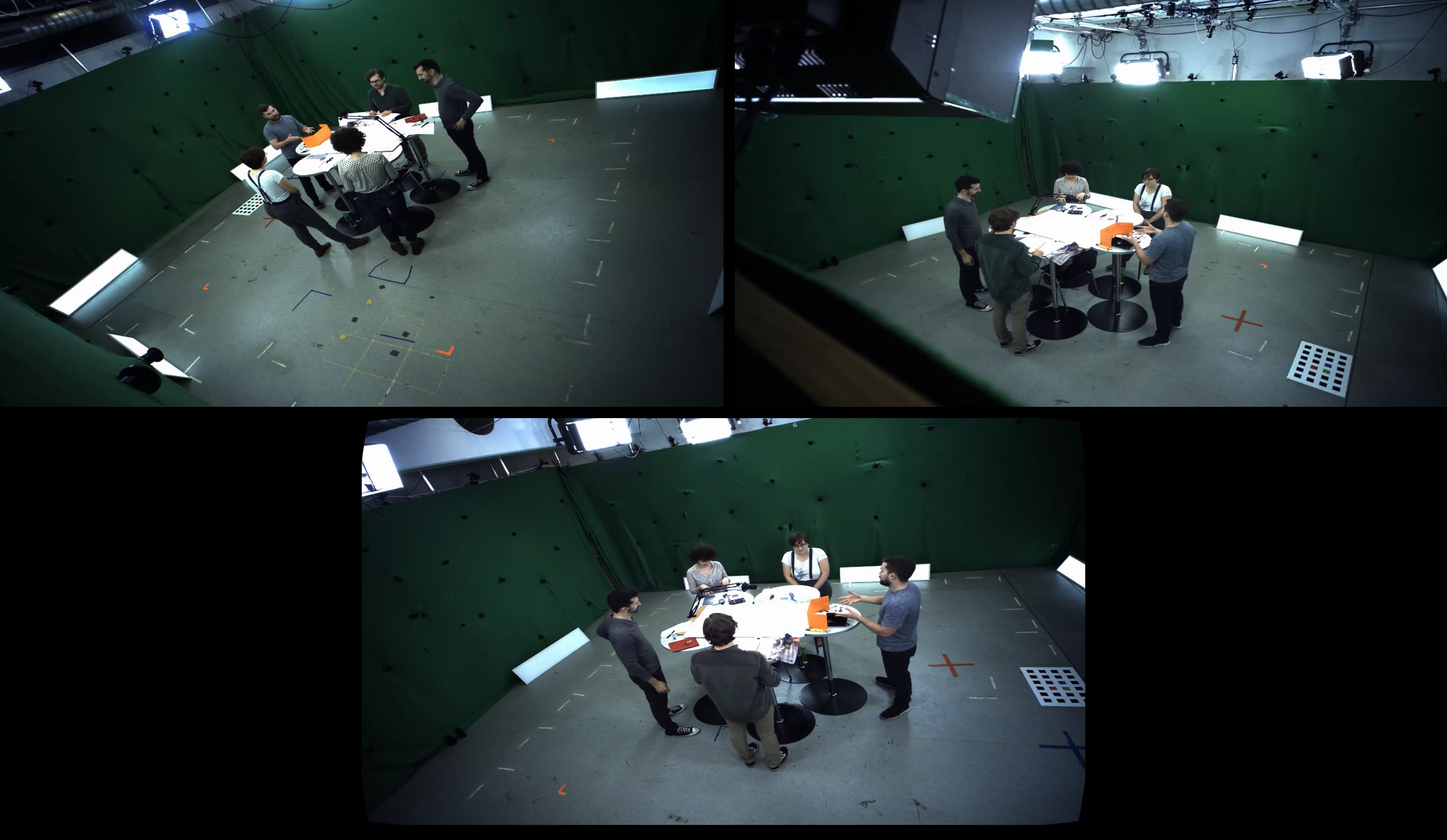}
        \caption{Raw video of the interaction.}
        \label{fig:sample-data-raw}
     \end{subfigure}
     \hfill
     \begin{subfigure}[t]{0.48\linewidth}
         \centering
         \includegraphics[width=\textwidth]{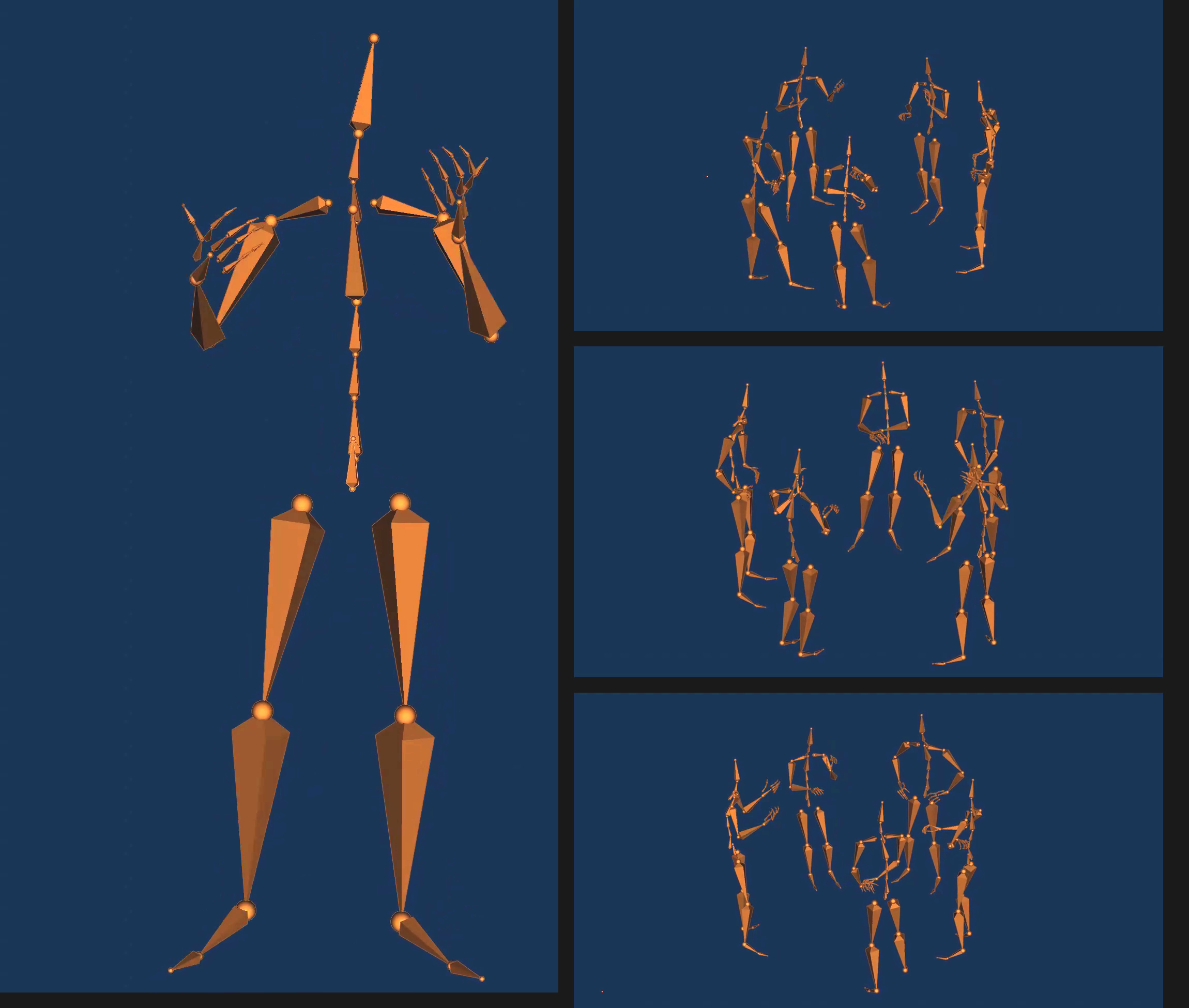}
         \caption{Skeletal representation.}
         \label{fig:sample-data}
     \end{subfigure}
     \hfill
     \caption{Example snapshot of thin-slices from the DnD Group Gesture dataset~\citep{Mughal2024-pi} shown to users in the perception study. (a) Video data from multiple cameras, and (b) the skeletal representation of the group and a single member.}
     \label{fig:sample-data-vid-skel}
     \vspace{-0.3cm}
\end{figure}

For our analysis, we require datasets that capture authentic human social interactions in multiparty settings. We selected the DnD Group Gesture dataset~\citep{Mughal2024-pi}, which records groups of five participants engaged in ``Dungeons and Dragons'' (DnD) gameplay sessions. This dataset is particularly valuable as it captures naturalistic social dynamics occurring within a structured collaborative activity, providing rich examples of spontaneous multimodal communication behaviours. The dataset provides RGB videos and BVH\footnote{\url{https://research.cs.wisc.edu/graphics/Courses/cs-838-1999/Jeff/BVH.html}} (Biovision Hierarchy) files that contain both skeletal structure information and the corresponding motion capture data recorded during gameplay. Each BVH file encodes a hierarchical representation of joint positions and rotations, allowing for precise tracking of body movements across multiple participants simultaneously. This skeletal representation offers several methodological advantages over RGB video data for our research objectives. The dataset also provides us with the audio files associated with each participant in that session of DnD.

First, skeletal data (\Cref{fig:sample-data}) enables us to perform controlled interventions on specific motion parameters while preserving the overall structural integrity of the movements. 
We can systematically manipulate gesture timing, amplitude, or coordination patterns between participants\textemdash interventions (\Cref{ssec:interventions}) that would be prohibitively complex or visually inconsistent if attempted on RGB video. Second, the skeletal format provides a dimensionality reduction that focuses our analysis specifically on movement dynamics rather than potentially confounding visual elements such as clothing, lighting, or facial expressions which is important for our chosen subjective measures (\Cref{ssec:subjective-measures}). Finally, this representation facilitates quantitative analysis of spatio-temporal coordination patterns between multiple participants, which is central to our investigation of social signals in group settings.

Our pipeline processes BVH files to extract movement features relevant to social coordination, applies systematic interventions to these features, and then utilises the modified skeletal animations for subsequent analysis.
We use thin-slices~\citep{Murphy2021-dn}, which requires us to slice our movement and audio features into \(30\)-second slices yielding \(\approx 145\) slices, upon which we apply interventions in order to test our measures.
This approach allows us to establish causal relationships between specific movement or audio parameters and perceived social dynamics within multiparty interactions.

\subsection{Motivation for Subjective Measures}\label{ssec:subjective-measures}

To assess perceptual differences between ground truth video recordings of participants in the DnD Dataset~\citep{Mughal2024-pi} to their skeletal embodiment derived from the BVH-based motion capture, we conduct a perception study. The experimental design employed the Perceived Conversation Quality framework~\citep{Raman2023-du} to measure conversation quality factors, alongside a modified Artificial Social Agent Questionnaire~\citep{Fitrianie2022-bj} to help quantify the \textit{perceived} human likeness of the agents. This methodological approach allows us to control for and account for potential perceptual shifts occurring when transitioning from video recordings to skeletal representations.

Perceived Conversation Quality metrics provide quantifiable measures for interpersonal relationships, the nature of interaction, and equality of opportunity that is also expected to capture engagement~\citep{Raman2023-du}. We selected this framework specifically because conversational dynamics are primarily conveyed through motion patterns and interactional timing, elements that should theoretically be preserved in skeletal representations despite the reduction in visual information. 
Human likeness assessment, as operationalized by the Artificial Social Agent Questionnaire, complements these metrics by evaluating how anthropomorphic features are perceived across different embodiment conditions~\citep{Fitrianie2022-bj}. This dual measurement approach allows us to examine both the quality of the observed interactions and the extent to which the skeletal representations preserve human-like qualities compared to the video recordings.
This dimension is particularly relevant as skeletal representations strip away surface-level visual cues while preserving motion dynamics, creating a controlled experimental condition to isolate movement-based factors in human perception, providing insight into which elements of human communication remain robust across different representation modalities. It is important to note that we are interested in the perception (by an observer) of conversation quality and human likeness rather than the anecdotal experience of the participants themselves in terms of likeability, sociality etc., which will be associated with in-group cues like clothing. This further motivates our choice of instruments and skeleton-representation for this exploratory study.

Our analysis focuses on the aspects of human behaviour that pertain to interpersonal synchrony, individual synchrony patterns (intra-personal) and distributional characteristics - taken together these aspects provide a window into how human behaviour is holistically perceived by others~\citep{Kendon1970-yw,Kendon1980-qo,Tomashin2022-wj,Alderisio2017-se,Trujillo2023-yz}. While interpersonal synchrony captures the perceived dynamic, moment-to-moment coordination between individuals, individual patterns reveal how observers interpret consistent tendencies and unique characteristics of each person. Distributional characteristics, in turn, provide insight into how broader patterns of behaviour vary across people and groups. The results of this study can be found in \Cref{ssec:exp-control} and more details about the setup are available in the Appendix.

\subsection{Interventions}\label{ssec:interventions}
Our study employs a series of targeted interventions designed to manipulate specific aspects of behaviour in thin slices of group interaction. These interventions allow us to examine how subtle changes in behavioural elements affect our chosen metrics in  \Cref{ssec:behavioural-measures}.
The interventions broadly target: interpersonal synchrony, self-synchrony, and distributional characteristics.

\paragraph{Movement Dampening}
Gesture kinematics have been shown to significantly influence social perception. Prior studies indicate that the speed of gestural movements can modulate perceived personality traits such as extraversion~\citep{Neff2010-mi, Niu2022-mn}, while the acceleration profiles of gestures impact how well gestures are perceived to match accompanying speech~\citep{Machiel2013-cj, Ruhlemann2024-oh, Cook2013-ul, Saerbeck2010-gs}. Additionally, larger hand movement amplitudes have been associated with communicative intent~\citep{Ruhlemann2024-oh, Deshmukh2018-ut,Pelachaud2009-qj}, suggesting that more expansive gestures convey greater social engagement.~\citet{Deshmukh2018-ut} demonstrated that increases in both the speed and amplitude of gesturing lead to higher ratings on the Godspeed questionnaire~\citep{Bartneck2023-nq}, particularly along dimensions of perceived anthropomorphism and likeability. Beyond these effects, recent work also shows that visual bodily signals, particularly from the upper body, play a crucial role in allowing observers to anticipate conversational turn-taking events~\citep{Ter-Bekke2024-sh}.
However, while these findings robustly link gesture kinematics to social perceptions, they often stop short of specifying the exact quantitative functions that help measure these effects. To systematically explore this space, we introduce a targeted manipulation of gesture kinematics by  dampening the movement of the hands and arms in the skeletal embodiment. By controlling movement magnitude in this way, we aim to causally assess the impact of gesture intensity on multimodal coordination metrics.
We hypothesize that altering gesture kinematics will lead to measurable changes at both the intra and inter personal levels. Specifically, we expect this intervention to influence joint-level Recurrence Quantification Analysis (RQA) within individuals, Cross-Recurrence Quantification Analysis (CRQA) between individuals, Soft Dynamic Time Warping (Soft-DTW) distances between gesture trajectories across individuals, and the Multiscale Beat Consistency between gestures and speech within an individual and between individuals. 
We dampen the movements of hands in the motion capture through the use of a Gaussian filter, which effectively functions as a low pass filter and thereby suppresses or dampens some of the hand movements and gestures. 
We apply this one level up the kinematic chain in order to ensure that the dampening occurs in the hands themselves. This is required due to the way motion is represented in BVH files. 

\paragraph{Speech-Gesture Delay}

We introduce delays in speech onset, which could also be seen as delays in response times between individuals, to disrupt the natural temporal alignment between gestures and speech. Prior work by~\citet{Ter_Bekke2024-uy} found that, on average, gestures preceded their semantically corresponding words by approximately $0.724~(\sigma=\pm 0.730)$ seconds during free-form conversation, providing a useful estimate for the typical temporal lag between modalities. Additionally, they reported shorter-range dependencies between gesture strokes and prosodically accented speech events, such as pitch accents, indicating that multiple levels of fine-grained coordination exist in natural communication. For the purposes of our intervention, we focus on the longer-range dependency observed during free-form conversation, as it better reflects the spontaneous, multiparty social settings captured in the DnD Group Gesture dataset~\citep{Mughal2024-pi}. Due to the impracticality of manually identifying and adjusting individual gesture-speech alignments for each participant, we instead apply a uniform delay of \(0.724 + \sigma\), across each individual's entire audio track. While this approach sacrifices fine-grained specificity, it provides a systematic perturbation that allows us to examine whether disrupting speech-gesture synchrony globally affects perceived social coordination. By introducing this misalignment, we hypothesize that self and interpersonal beat consistency will be degraded.

\paragraph{Voice Pitch Variance Reduction}

Previous research has demonstrated that vocal information plays a crucial role in shaping listeners' perceptions of speaker traits such as confidence and social dominance~\citep{Guyer2019-hc, Guyer2021-gq}. In particular,~\citet{Guyer2019-hc} showed that manipulating a speaker’s pitch can significantly alter perceived confidence levels: lowering the fundamental frequency (F0) by $20$ Hz led to higher confidence ratings, while raising it by $120$ Hz had the opposite effect. These findings highlight that even relatively small perturbations in prosodic features can meaningfully influence social evaluations.
Motivated by these insights, we introduce a pitch variance reduction intervention in our analysis. Specifically, we constrain the F0 trajectories of speakers within a limited range around their mean F0, effectively reducing prosodic variability without altering the overall verbal content. This manipulation enables us to isolate the role of vocal expressivity in multimodal social coordination.
To quantify the impact of pitch variance reduction, we compute the Soft-DTW distance between the original and altered F0 streams for each thin-slice segment. Soft-DTW is particularly well-suited for this task as it allows for flexible, differentiable alignment between sequences, making it sensitive to subtle changes in prosodic contours. We hypothesize that reducing pitch variability will disrupt natural prosodic dynamics, leading to measurable changes in Soft-DTW distances.

\section{Experiments}\label{sec:experiments} 


\begin{figure}[t]
     \centering
     \begin{subfigure}[b]{0.48\linewidth}
         \centering
         \includegraphics[width=\textwidth]{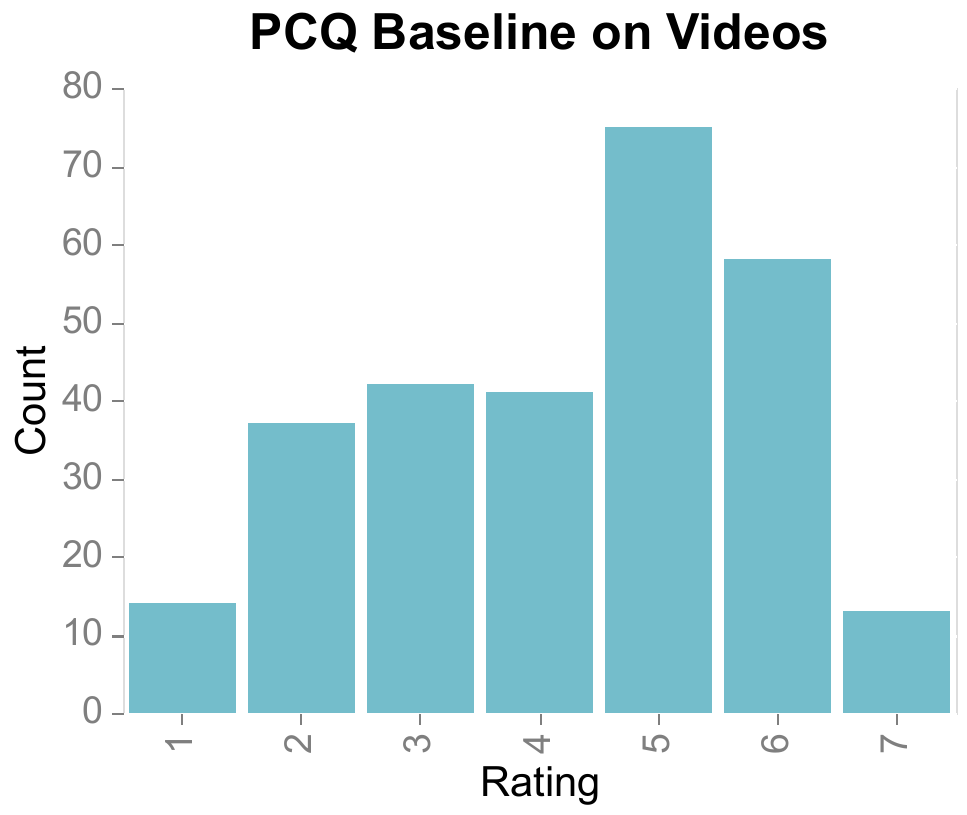}
         \caption{PCQ ratings on the baseline videos with no interventions.}
         \label{sfig:pcq-baseline}
     \end{subfigure}
     \hfill
     \begin{subfigure}[b]{0.48\linewidth}
         \centering
         \includegraphics[width=\textwidth]{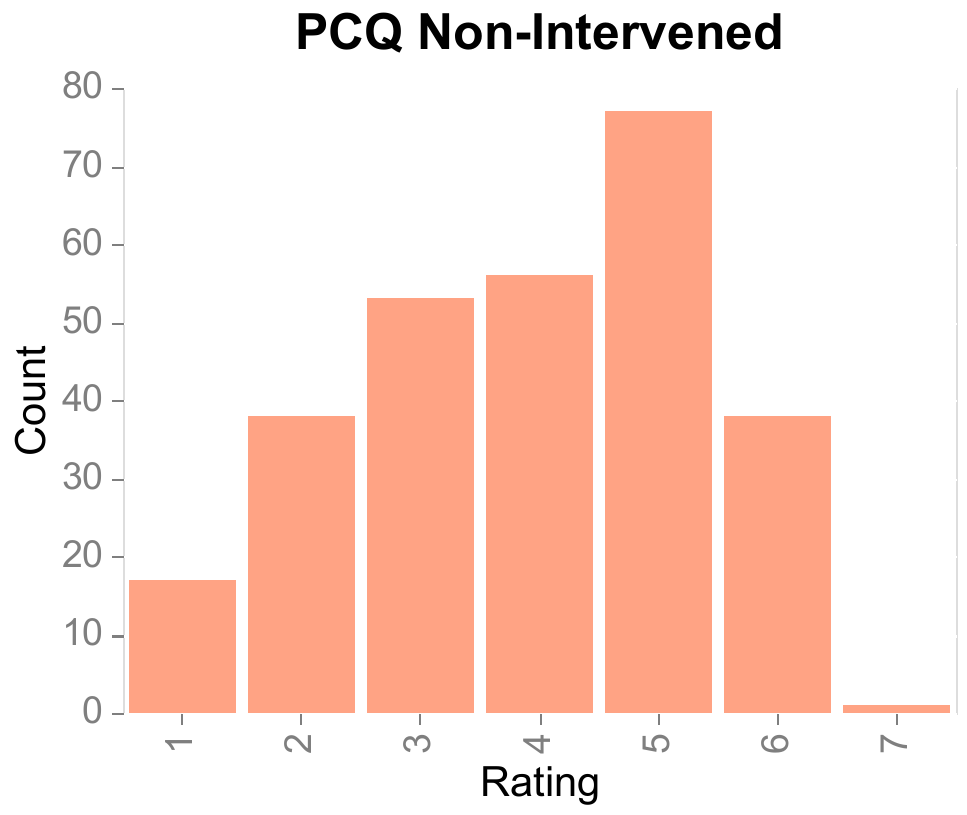}
         \caption{PCQ ratings on the non-intervened stick representations.}
         \label{sfig:pcq-stick-hist}
     \end{subfigure}
     \hfill
     \begin{subfigure}[b]{0.48\linewidth}
         \centering
         \includegraphics[width=\textwidth]{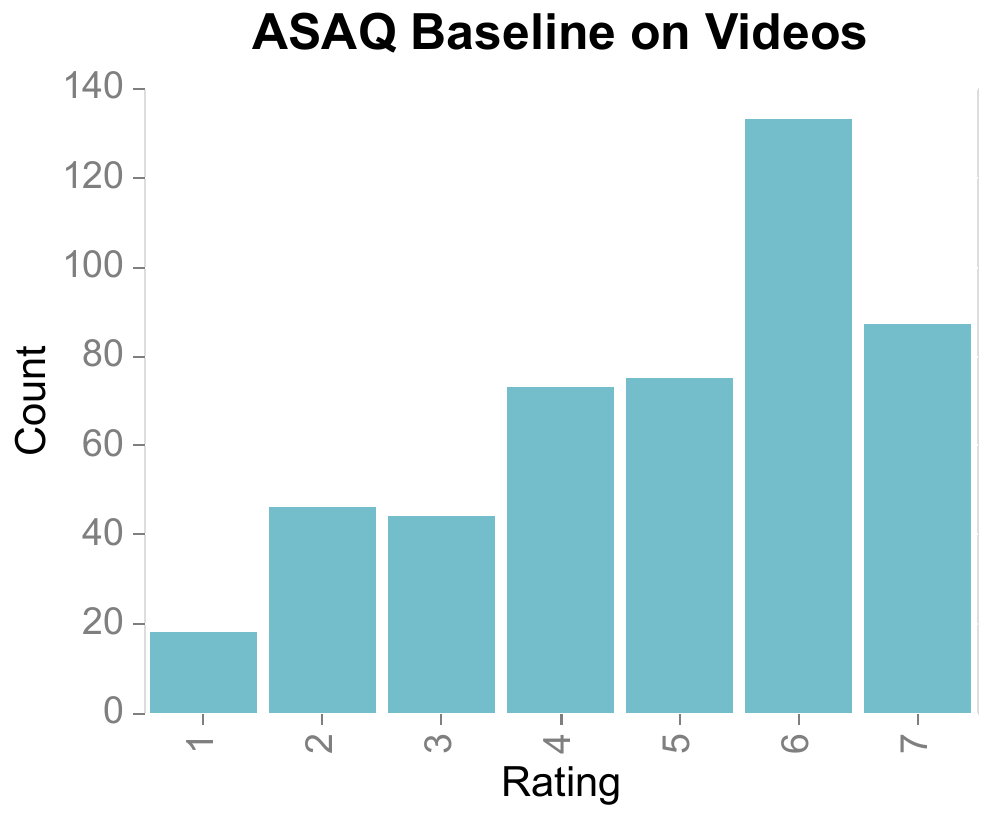}
         \caption{ASAQ ratings on the baseline videos with no interventions.}\label{sfig:asaq-baseline-hist}
     \end{subfigure}
     \hfill
     \begin{subfigure}[b]{0.48\linewidth}
         \centering
         \includegraphics[width=\textwidth]{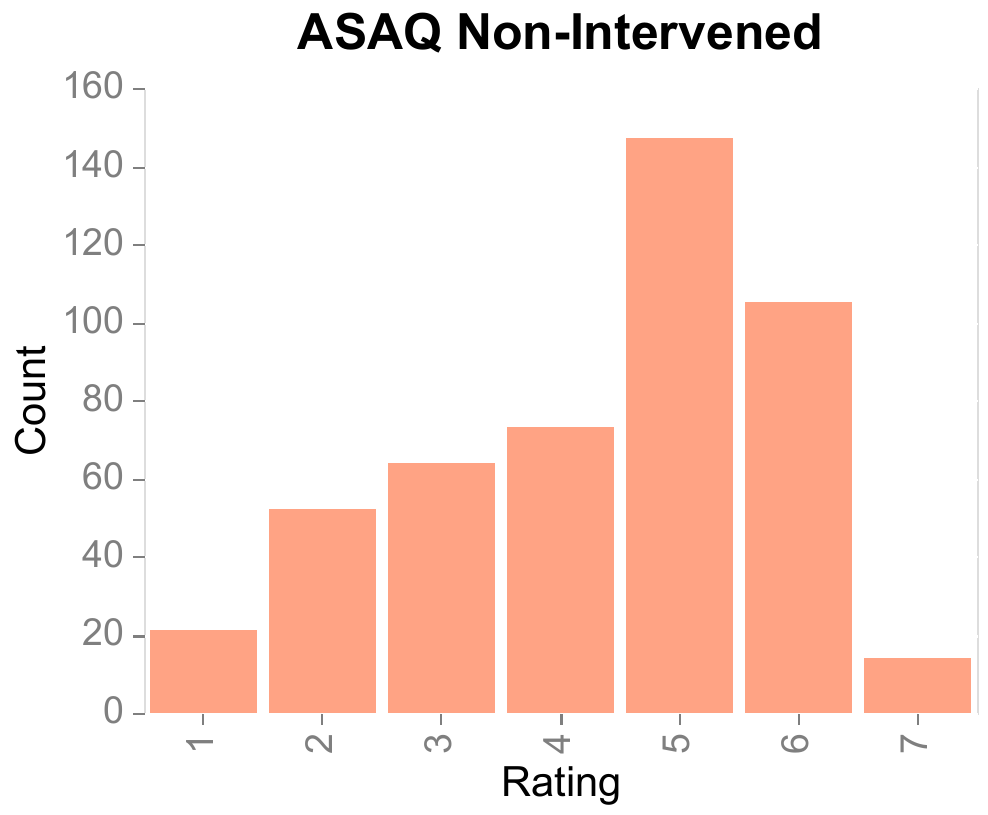}
         \Description[ASAQ NI Histogram.]{ASAQ Non-Intervened Histogram.}
         \caption{ASAQ ratings on the non-intervened stick representations.}\label{sfig:asaq-stick-base-hist}
     \end{subfigure}
     \hfill
        \caption{PCQ and ASAQ ratings from a control experiment between the baseline videos and the stick representations of humans from the video.  The control experiment is done to show the possible effect of changing the representation from videos to the stick skeleton.}
        \label{fig:asaq-pcq-baselines}
\end{figure}


\begin{table}[t]
\caption{LMEM results predicting CRQA \%DET. Dampening strength at \(10\). Complete table in Appendix.}
\label{tab:crqa-lmem-det-dampx10}
\small
\centering
{\tabcolsep=2pt\def\arraystretch{1.0}
\begin{tabularx}{\linewidth}{@{}>{\arraybackslash}l *5{Y}}
\toprule
\textbf{Predictor} & \textbf{Coef.} & \textbf{p-value} & \multicolumn{2}{c}{\(\mathbf{95\%}\) \textbf{CI}} \\
\midrule
Intercept                                           & -0.093 &     0.000  & -0.134 & -0.052  \\
Condition: Dampened                                   &  0.115 &     0.000  &  0.073 &  0.157  \\
Group Variance                                      &  0.301 &            &        &         \\
\bottomrule
\end{tabularx}}

\end{table}

\begin{table}[t]
\caption{LMEM results predicting CRQA MeanLR. Dampening strength at \(10\). Complete table in Appendix.}
\label{tab:crqa-lmem-meanlen-dampx10-small}
\small
\centering
{\tabcolsep=2pt\def\arraystretch{1.0}
\begin{tabularx}{\linewidth}{@{}>{\arraybackslash}l *5{Y}}
\toprule
\textbf{Predictor} & \textbf{Coef.} & \textbf{p-value} & \multicolumn{2}{c}{\(\mathbf{95\%}\) \textbf{CI}} \\
\midrule
Intercept                                           & -0.169 &     0.000  & -0.210 & -0.128  \\
Condition: Dampened                                   &  0.451 &     0.000  &  0.410 &  0.492  \\
Group Variance                                      &  0.334 &            &        &         \\
\bottomrule
\end{tabularx}}

\end{table}

\begin{table}[htbp]
\caption{LMEM predicting inter-person Soft-DTW distances. Dampening strength at \(10\). Complete table in Appendix.}
\label{tab:sdtw-inter-person-intervenedx10-small}
\small
\centering
{\tabcolsep=2pt\def\arraystretch{1.0}
\begin{tabularx}{\linewidth}{@{}>{\arraybackslash}l *5{Y}}
\toprule
\textbf{Predictor} & \textbf{Coef.} & \textbf{p-value} & \multicolumn{2}{c}{\(\mathbf{95\%}\) \textbf{CI}} \\
\midrule
Intercept                                           &  0.323 &     0.000  &  0.272 &  0.373  \\
Condition: Intervened                               & -0.637 &     0.000  & -0.688 & -0.586  \\

Group Variance                                      &  0.376 &            &        &         \\
\bottomrule
\end{tabularx}}
\end{table}



\begin{figure}[t]
     \centering
     \begin{subfigure}[t]{0.48\linewidth}
         \centering
         \includegraphics[width=\textwidth]{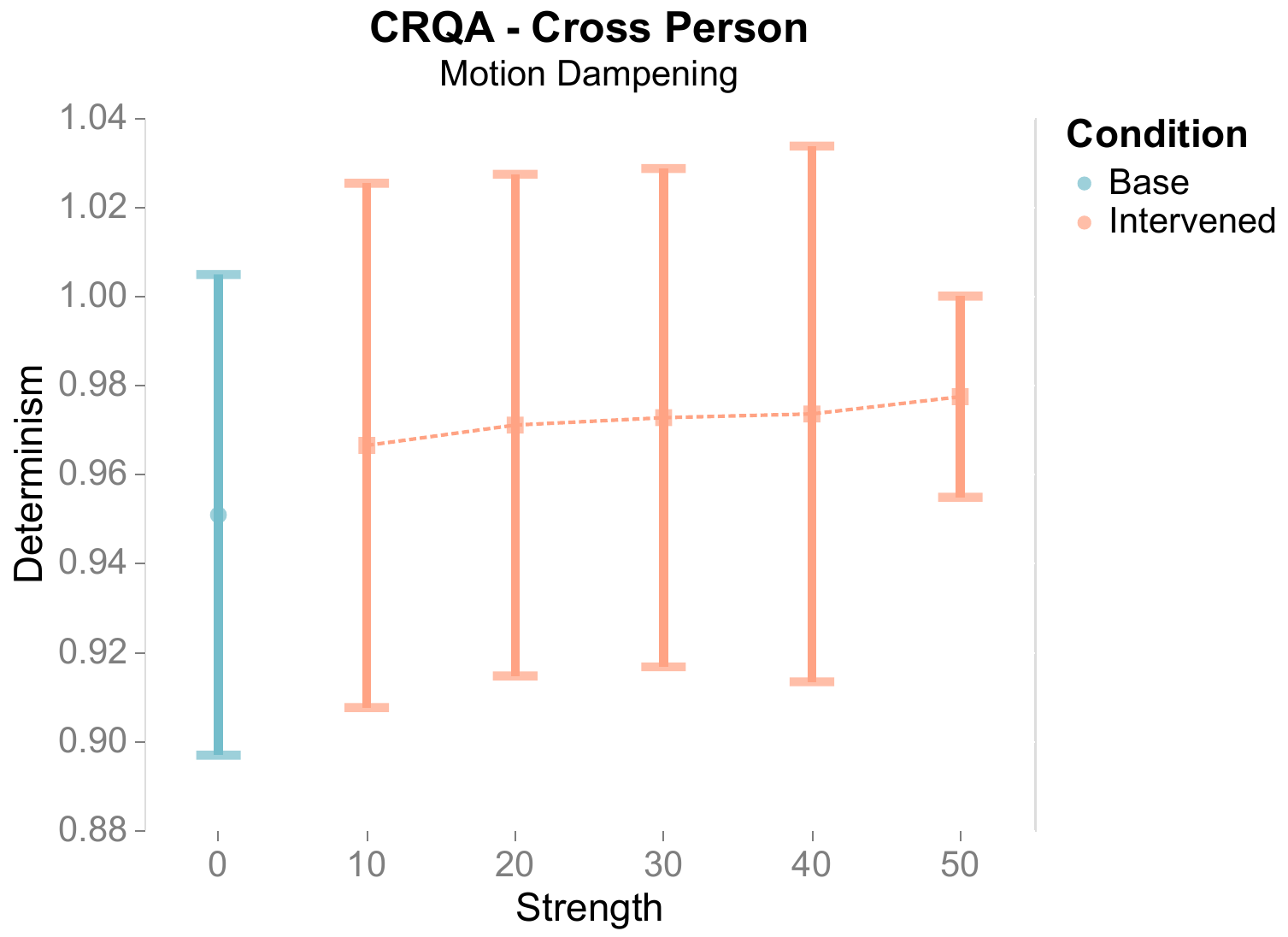}
         \caption{CRQA \%DET of an individual averaged against all other conversation partners.}
         \label{fig:cross-det-err-plt}
     \end{subfigure}
     \hfill
     \begin{subfigure}[t]{0.48\linewidth}
         \centering
         \includegraphics[width=\textwidth]{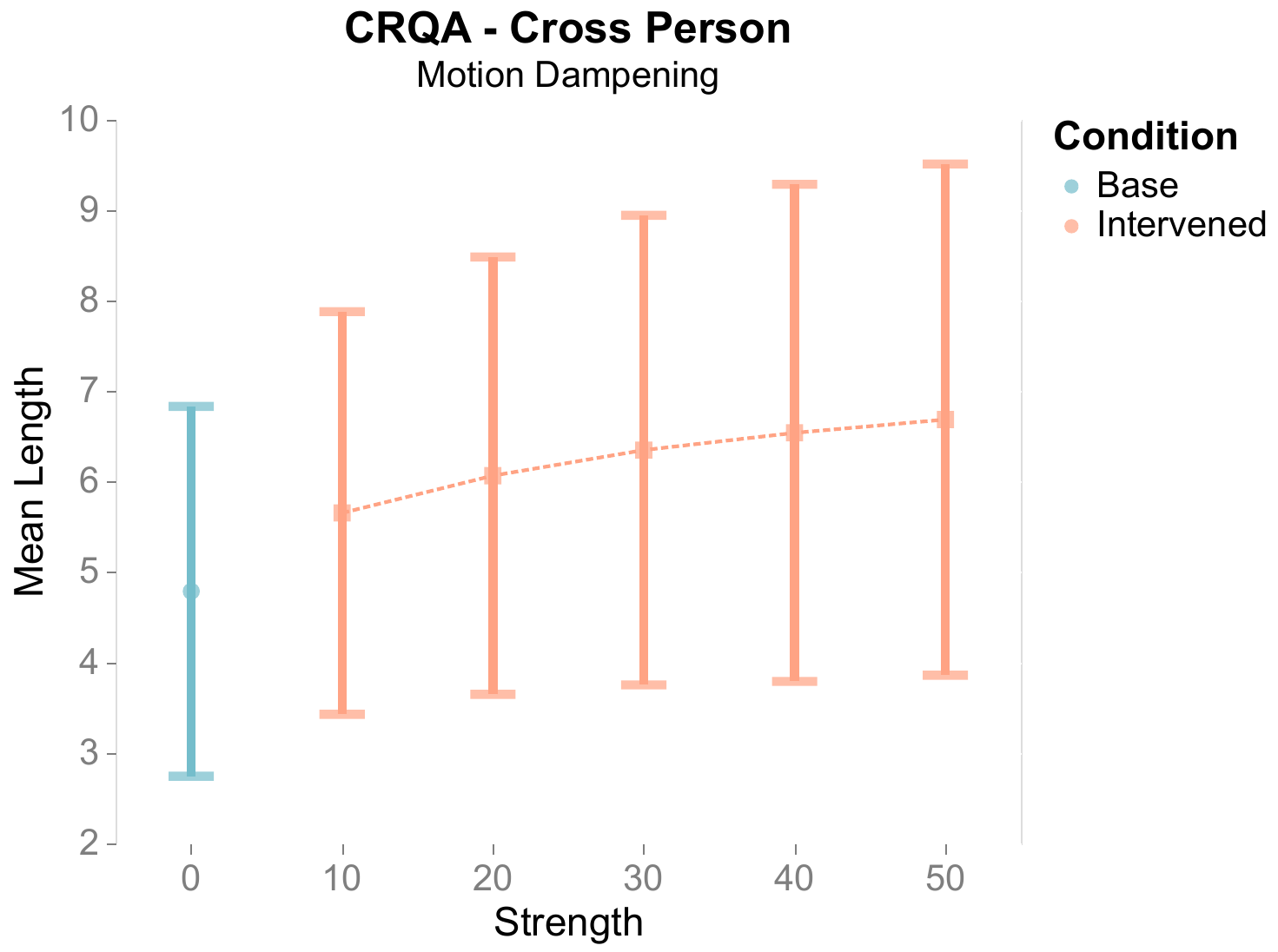}
         \caption{CRQA MeanLR of an individual averaged against all other conversation partners.}
         \label{fig:cross-meanl-err-plt}
     \end{subfigure}
     \hfill
     \caption{Figure shows the CRQA Average \%DET and MeanLR in the gesture signal of each individual. The dampened motion leads to more predictable motion.}
     \label{fig:crqa-all-err-plts}
     \vspace{-0.3cm}
\end{figure}

\begin{figure}[t]
     \centering
     \begin{subfigure}[t]{0.48\linewidth}
         \centering
         \includegraphics[width=\textwidth]{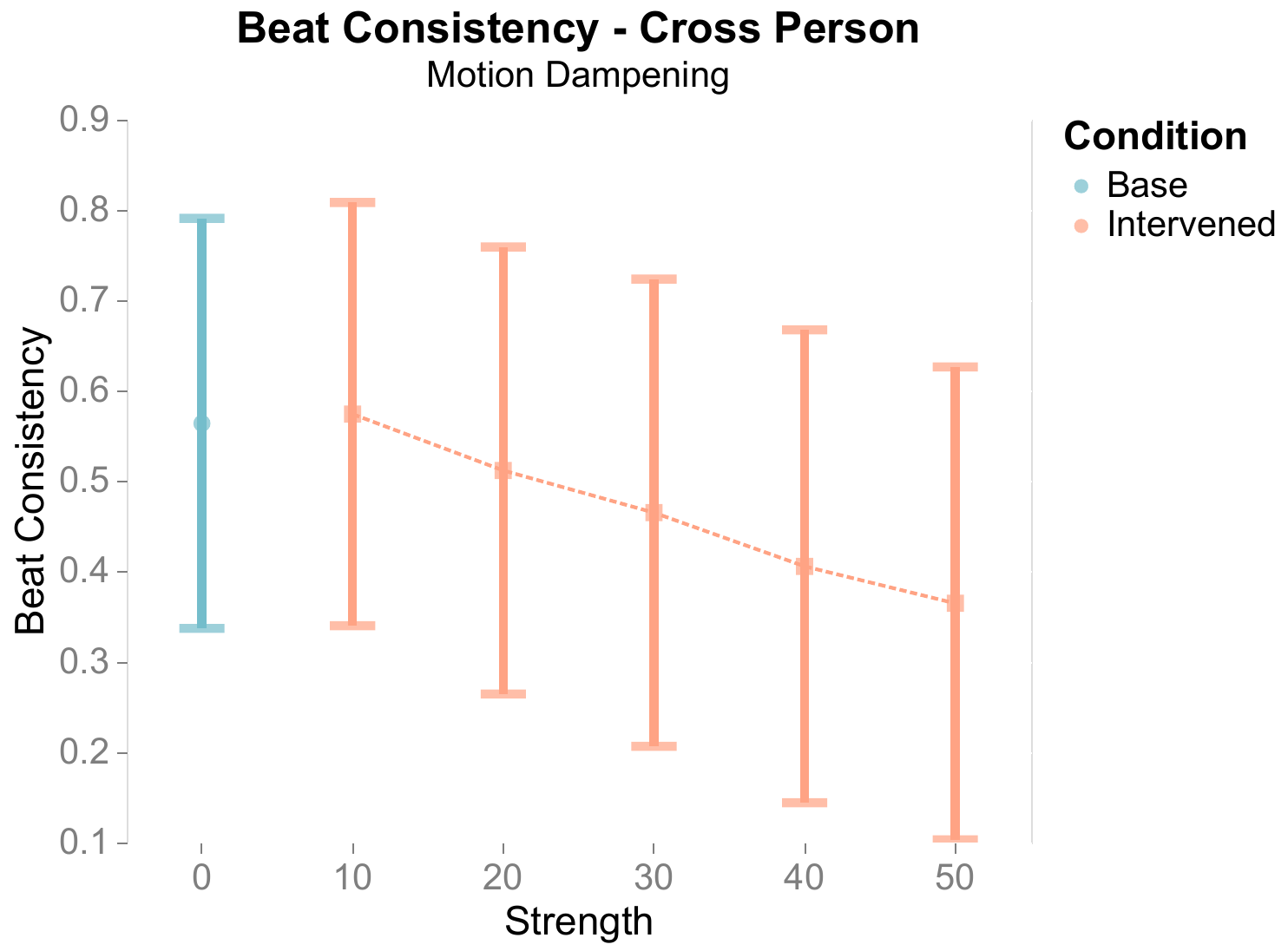}
         \caption{Beat Consistency of an individual averaged against all others when gestures are dampened.}
         \label{fig:cross-bc-damp-err-plt}
     \end{subfigure}
     \hfill
     \begin{subfigure}[t]{0.48\linewidth}
         \centering
         \includegraphics[width=\textwidth]{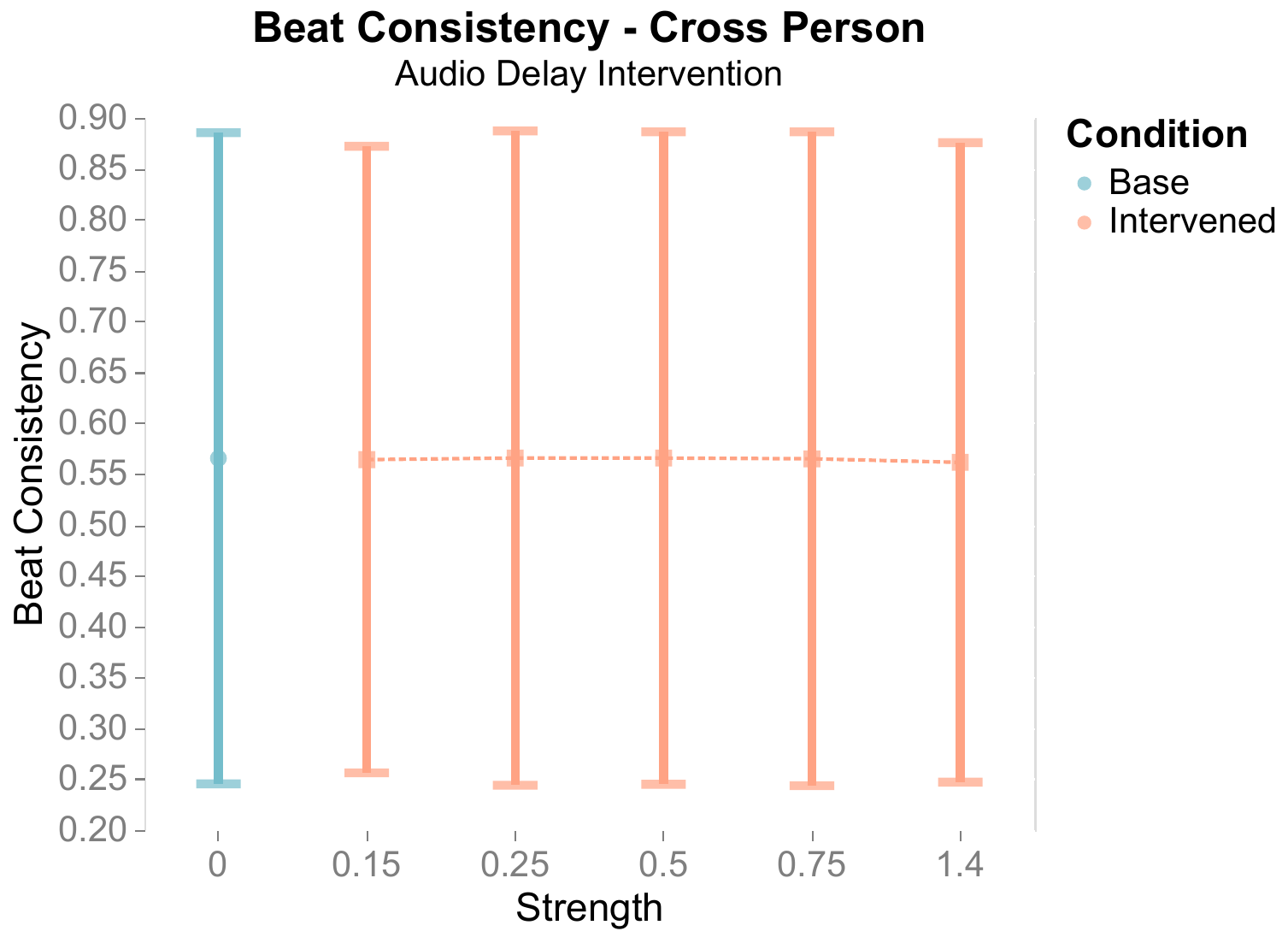}
         \caption{Beat Consistency of an individual averaged against all others with gesture onset delay.}
         \label{fig:cross-bc-delay-err-plt}
     \end{subfigure}
     \hfill
     \caption{Figure shows the cross-modal inter-person beat consistency. Averaged against all conversation partners.}
     \label{fig:cross-bc-plots}
\end{figure}



\subsection{Exploratory Perception Study}\label{ssec:exp-control}
In order to quantify the change in perception that is caused by changing representations (video $\rightarrow$ stick skeletons), we presented thirteen \(30\)-second thin-slices to \(27\) participants. Each clip was shown to an average of \(\sim 3.0\) participants who filled out the ASAQ~\citep{Fitrianie2022-bj} and PCQ~\citep{Raman2023-du} instruments. \Cref{fig:asaq-pcq-baselines} shows the perceived changes between the video and stick skeleton representations as ratings received on the ASAQ and PCQ instruments. There is a change in the ASAQ ratings indicating that subjects found that stick representations to be less ``human-like'' in appearance (Mean (video $-$ stick figure)  = 0.51, SD = 0.67, t(27) = 4.07, p $<$ .001). While the change in PCQ reflects lower quality of perceived conversations (Mean (video $-$ stick figure)  = 0.34, SD = 0.47, t(27) = 3.89, p $<$ .001). Both measures show a change likely due to the skeletal representations missing articulations not afforded in the representation such as facial expressions.

\subsection{Movement Dampening}\label{ssec:exp-move-damp}

\begin{figure}[t]
     \centering
     \begin{subfigure}[t]{0.48\linewidth}
         \centering
         \includegraphics[width=\textwidth]{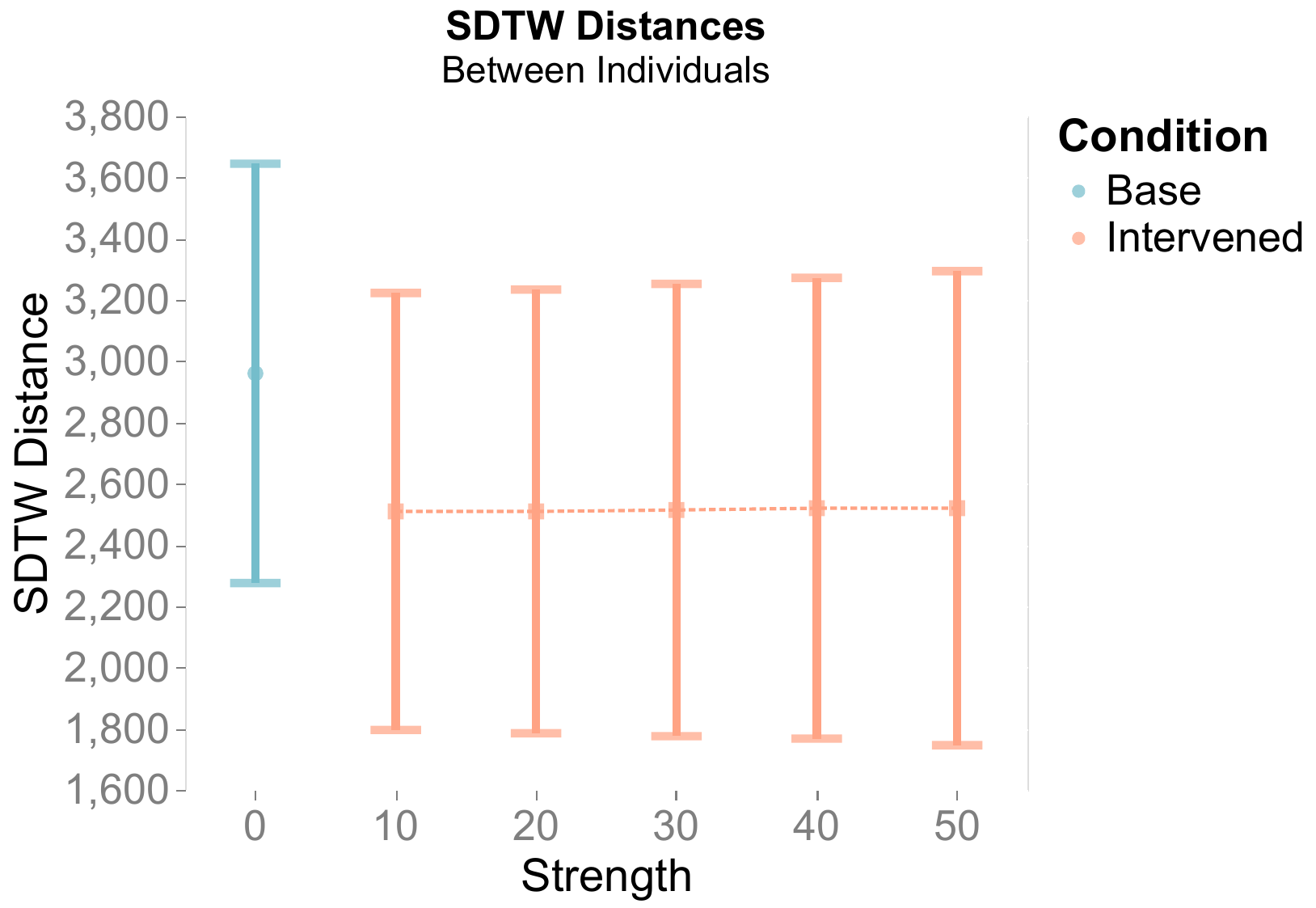}
        \caption{Inter-person SDTW distances between joints. Dampening signals looses variations and hence reduces the alignment cost recorded by SDTW.}
        \label{fig:cross-sdtw-mov-damp-err-plt}
     \end{subfigure}
     \hfill
     \begin{subfigure}[t]{0.48\linewidth}
         \centering
         \includegraphics[width=\textwidth]{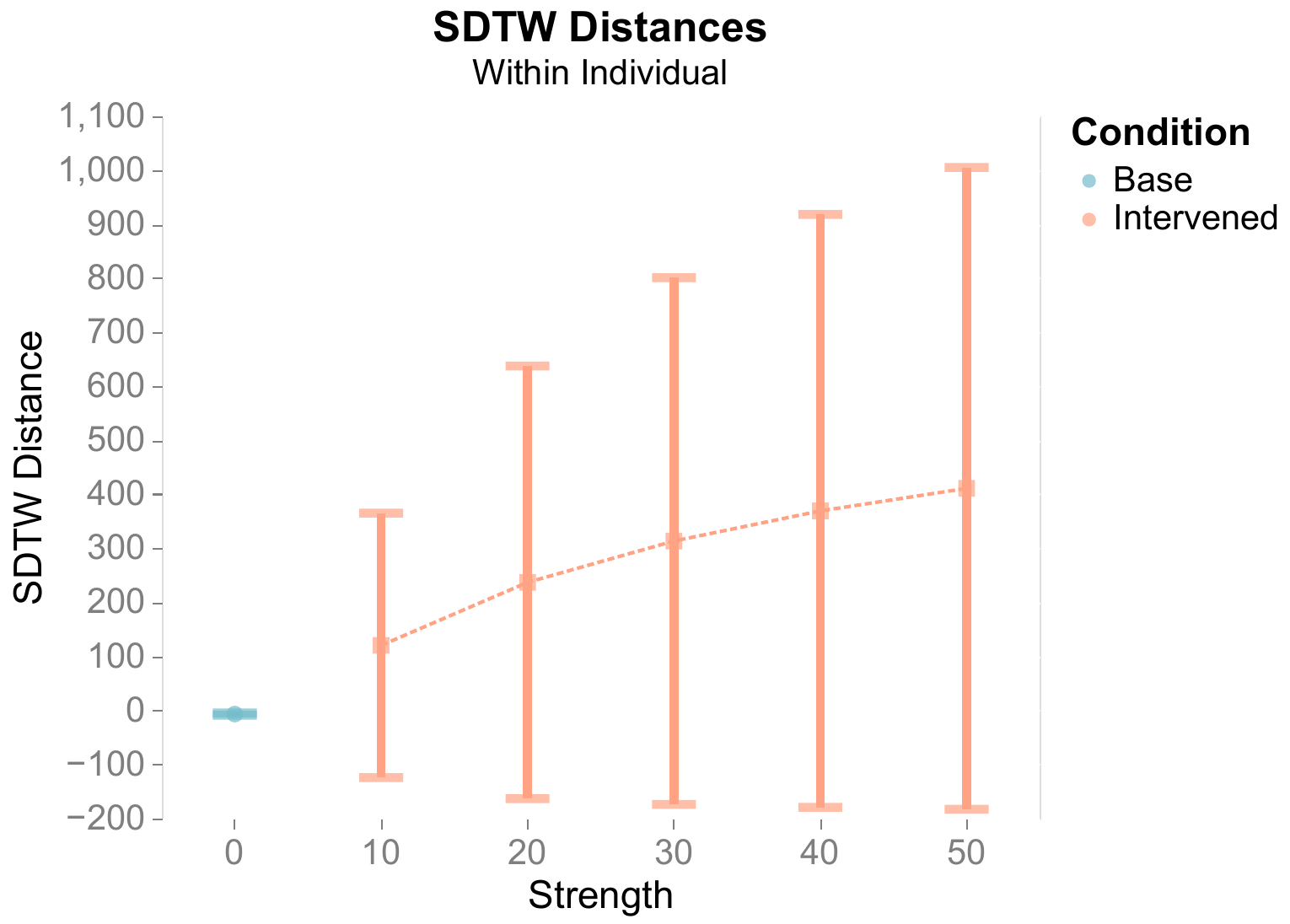}
         \caption{Intra-person SDTW distances of altered motion against unaltered ground truth.}
         \label{fig:ind-sdtw-mov-damp-err-plt}
     \end{subfigure}
     \hfill
     \caption{Figure shows SoftDTW distances calculated between ground-truth data and the manipulated variations.}
     \label{fig:softdtw-err-plots-both}
     \vspace{-0.2cm}
\end{figure}

For analysing the intra and inter person effects we rely on linear mixed-effects models (LMEM), with the implementation from statsmodels~\citep{Seabold2010-qf}. Here we mainly talk about the first strength level that yields a significant change in metrics, however all strength levels (\(\sigma \geq 10\)) yield a \(p < 0.05\). Detailed LMEM summary tables can be found in the Appendix.
It should be noted that, despite high variability in the  data, the linear mixed-effects model identifies a consistent effect after accounting for subject-level differences.
The LMEM shows that the Dampened condition raised \%DET for the Right-Hand by $0.0.026$ ($p < .001$). A similar increase was observed for the Left-Hand. Similarly, MeanLR for the Right-Hand increased by $1.56$ units in the Dampened condition ($p < .001$). Left-Hand is raised by $\approx 1.2$ units under intervention. Dampening also consistently lowers beat consistency for strength $\geq 20$ across all people with the mean ($0.54$) decreasing  by $0.06$. LMEM summary tables for the intra-person RQA and Beat Consistency are available in the Appendix.
For analysing the effect between pairs of people, we make use of CRQA (\Cref{fig:crqa-all-err-plts}), Beat Consistency (\Cref{fig:cross-bc-plots}) and SoftDTW (\Cref{fig:cross-sdtw-mov-damp-err-plt}).~\Cref{tab:crqa-lmem-det-dampx10-small} shows the CRQA \%DET is substantially increased for the hands. Introducing dampening raised \%DET for Right-Hand by $0.115$ ($p < .001$) and boosted Left-Hand by an even larger $0.241$ ($0.115 + 0.126$, $p<.001$). Similarly, dampening raised the Mean Length of the Right-Hand by $0.451$ units ($p < .001$) and, after accounting for a small interaction ($-0.059$), the Left-Hand by $\approx 0.392$ units. 
Beat Consistency was also lowered between all individuals compared to the baseline as shown in \Cref{fig:cross-bc-damp-err-plt}. The intervention increased SDTW distance: across all joints, the Dampened condition was $\approx 0.6$ units lower than non-intervened ($p < .001$) as shown in~\Cref{tab:sdtw-inter-person-intervenedx10-small}, indicating that the variability in movement has gone down. Thus, the intervention produced a consistent, joint-independent reduction in movement variability (due to smoothing), which was detected by Soft-DTW.


\begin{table}[t]
\caption{LMEM predicting inter-person Beat Consistency for motion dampening strength of $10$.}
\label{tab:cross-bc-dampened-lmemx10}
\small
\centering
{\tabcolsep=2pt\def\arraystretch{1.0}
\begin{tabularx}{\linewidth}{@{}>{\arraybackslash}l *5{Y}}
\toprule
\textbf{Predictor} & \textbf{Coef.} & \textbf{p-value} & \multicolumn{2}{c}{\(\mathbf{95\%}\) \textbf{CI}} \\
\midrule
Intercept                   & 0.559 & 0.000 & 0.546 & 0.571 \\
Condition: Intervened       & 0.013 & 0.030 & 0.001 & 0.025 \\
Group Variance              & 0.021 &       &       &       \\
\bottomrule
\end{tabularx}}
\end{table}

\begin{table}[htbp]
\caption{LMEM predicting inter-person Beat Consistency for the audio delay manipulation. Delay set to \(0.25\)s.}
\label{tab:lmem-cross-bc-audio-delayx0.25}
\small
\centering
{\tabcolsep=2pt\def\arraystretch{1.0}
\begin{tabularx}{\linewidth}{@{}>{\arraybackslash}l *5{Y}}
\toprule
\textbf{Predictor} & \textbf{Coef.} & \textbf{p-value} & \multicolumn{2}{c}{\(\mathbf{95\%}\) \textbf{CI}} \\
\midrule
Intercept             & 0.565 & 0.000 & 0.552 & 0.578 \\
Condition: Intervened & -0.004 & 0.041 & -0.008 & -0.000 \\
Group Variance        & 0.047 &       &        &       \\
\bottomrule
\end{tabularx}}

\end{table}

\subsection{Speech-Gesture Delays}\label{ssec:exp-speech-gest-delay}
We assess the intervention using our Beat Consistency measure (figures of the intra-person case made available in the Appendix). 
In the intra-person scenario dampened gestures yield slightly lower consistency scores, although an LMEM indicates the manipulation’s effect is not strong enough. This is consistent for all delay values (see Appendix for detailed LMEM tables) for the intra-person case.
In the cross-person scenario, against a baseline mean of \(0.565\) with a delay of $0.25$s, the intervention produces a modest decrease of \(0.004\)~($p<0.05$) units as shown in \Cref{tab:lmem-cross-bc-audio-delayx0.25}. The remaining strengths do not reach significance, potentially due to sparse and uneven onsets per slice. The complete set of LMEM summary tables is available in the Appendix.

\subsection{Vocal Pitch Variance}\label{ssec:exp-vocal-pitch-var}


When Soft-DTW is applied to two identical F0 contours, the optimal warping path hugs the main diagonal and the resulting distance is essentially zero; any deviation from that diagonal reflects temporal or spectral differences and increases the cost.  In our data, comparing each speaker’s unaltered contour with itself produced distances at (or numerically indistinguishable from) zero, whereas comparing the pitch-manipulated contour with its unaltered counterpart yielded markedly higher values, consistent with a substantial prosodic change. Because the baseline distribution is degenerate at zero and therefore violates the assumptions of linear-mixed modelling, we evaluated the effect with a paired Wilcoxon signed-rank test, which confirmed a reliable increase in Soft-DTW distance (\(W = 1732\), \(p < 0.001\)). An error plot and a summary table for various intervention strengths are available in the Appendix for the interested reader.

\section{Discussion and Conclusion}\label{sec:discussion}

Our three intervention families demonstrate that distinct, theoretically motivated metrics reveal complementary facets of multiparty social behaviour. Dampening the kinematics of the hands and arms produced more internally predictable gestures (RQA \%DET and MeanLR), increased inter-agent gestural synchrony (CRQA \%DET) and compressed spatial variability (Soft-DTW), yet simultaneously weakened speech-gesture coupling (Beat Consistency). Introducing a uniform $1.4$s audio delay only marginally affected self beat-alignment but reliably lowered cross-person Beat Consistency, indicating that temporal mis-alignment degrades group-level coordination before it is noticeable within an individual. Flattening prosodic pitch variance left motion untouched but sharply increased Soft-DTW distances between original and altered F0 contours, confirming the measure’s sensitivity to subtle prosodic changes. Across all manipulations the hands proved the most responsive modality: they drove the largest gains in predictability under dampening and showed the clearest correspondence in objective changes
, underscoring their central role in signalling social engagement.

We observe that dampening movement inflates coordination metrics—yielding higher CRQA \%Determinism and mean line lengths and lower DTW distances (e.g., \citep{Trujillo2023-vm,Fan2024-ts}). This follows from the mechanics of these measures since they track co-occurrence in temporal dynamics. Since relative stillness is a common listening behaviour it constitutes a stable attractor. By reducing a speaker's movement, their kinematics align more with listeners' idling. However there is a risk of conflating idle synchrony with genuine dynamic coordination, which will also relate differently with social perceptions. Our sanity check warns that metrics often taken to index rapport (e.g., \citep{Trujillo2023-vm,Fan2024-ts}) may simply reflect shared idle attractors or phases, since they merely reflect recurrence around any stable pattern. For example, \citet{Kodama2024-ku} found that people who cannot see each other exhibit stronger head-movement “coordination” (via CRQA max line length and recurrence rate) than those in view—our analysis shows that it is risky to conclude that not seeing each other during conversation enhances synchrony or coordination between head movements, as it may actually be rooted in reduced or dampened movements. Further, checks are therefore needed to differentiate between dynamic coordination versus co-occurence of idling moments in conversation.

Overall, our results show that no single metric can fully assess social believability. Instead, a small suite of measures: dynamical structure via RQA/CRQA, cross-modal timing via Beat Consistency, and distributional similarity via Soft-DTW—provides complementary, diagnostic insights. Because each metric is sensitive to the specific property it targets, behaviour generation systems can treat them as a “palette”. For example, impose a minimum \%DET to preserve gestural structure, enforce beat-alignment thresholds to ensure audio-motion coherence, or monitor Soft-DTW to prevent over-smoothed motion. The consistent cross-group effects and minimal random-intercept variance in our models also suggest these measures are robust to individual differences, making them well suited for large-scale automated evaluation. Future work should incorporate head-pose and facial recurrence metrics to close the remaining gap in perceived realism, and explore whether integrating this metric suite into the training loop can steer generators toward truly socially coherent digital humans.

\section{Safe and Responsible Innovation Statement}\label{sec:safety-statement}



Digital humans and socially interactive agents promise transformative advances in education and healthcare, yet also pose ethical and societal challenges. Here, we link objective, low-level measures of multi-party social behaviour to high-level subjective perceptions—cautiously warning against any manipulative use of inferred human responses. To minimize methodological bias and foster transparency, we applied diverse interventional perturbations and ran an exploratory perception study on our motion representations. Finally, to ensure privacy and GDPR compliance, we used anonymized skeletal data (stripping faces, clothing, and backgrounds) under approvals from our university's Data Steward and Human Research Ethics Committee.


\bibliographystyle{ACM-Reference-Format}
\bibliography{paperpile}
\clearpage

\appendix

\section{Intra-person RQA}\label{ref:intra-person-rqa}

\begin{figure}[t]
     \centering
     \begin{subfigure}[b]{0.48\linewidth}
         \centering
         \includegraphics[width=\textwidth]{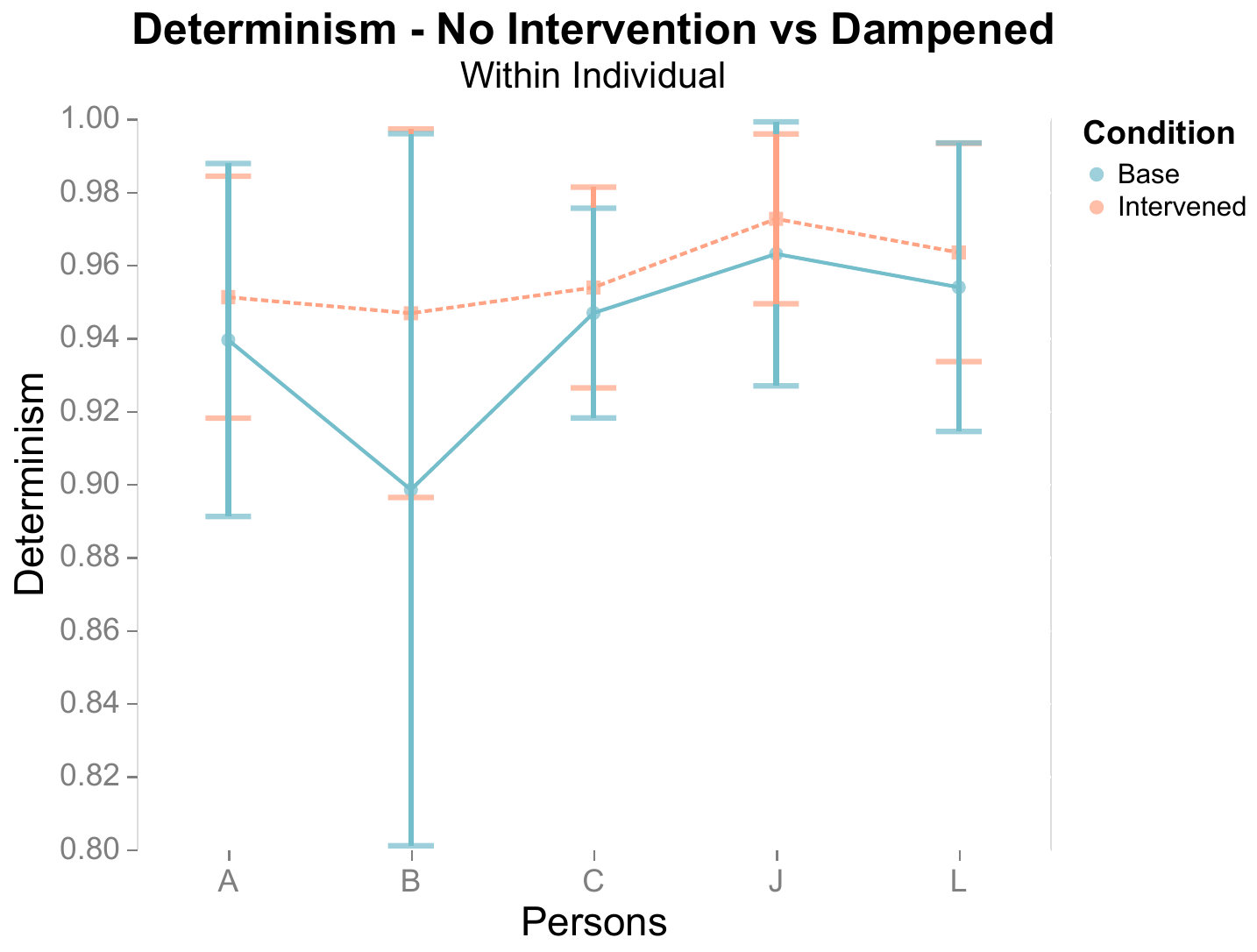}
         \caption{RQA Average \%DET in an individual.}
         \label{fig:indiv-det-err-plt}
     \end{subfigure}
     \hfill
     \begin{subfigure}[b]{0.48\linewidth}
         \centering
         \includegraphics[width=\textwidth]{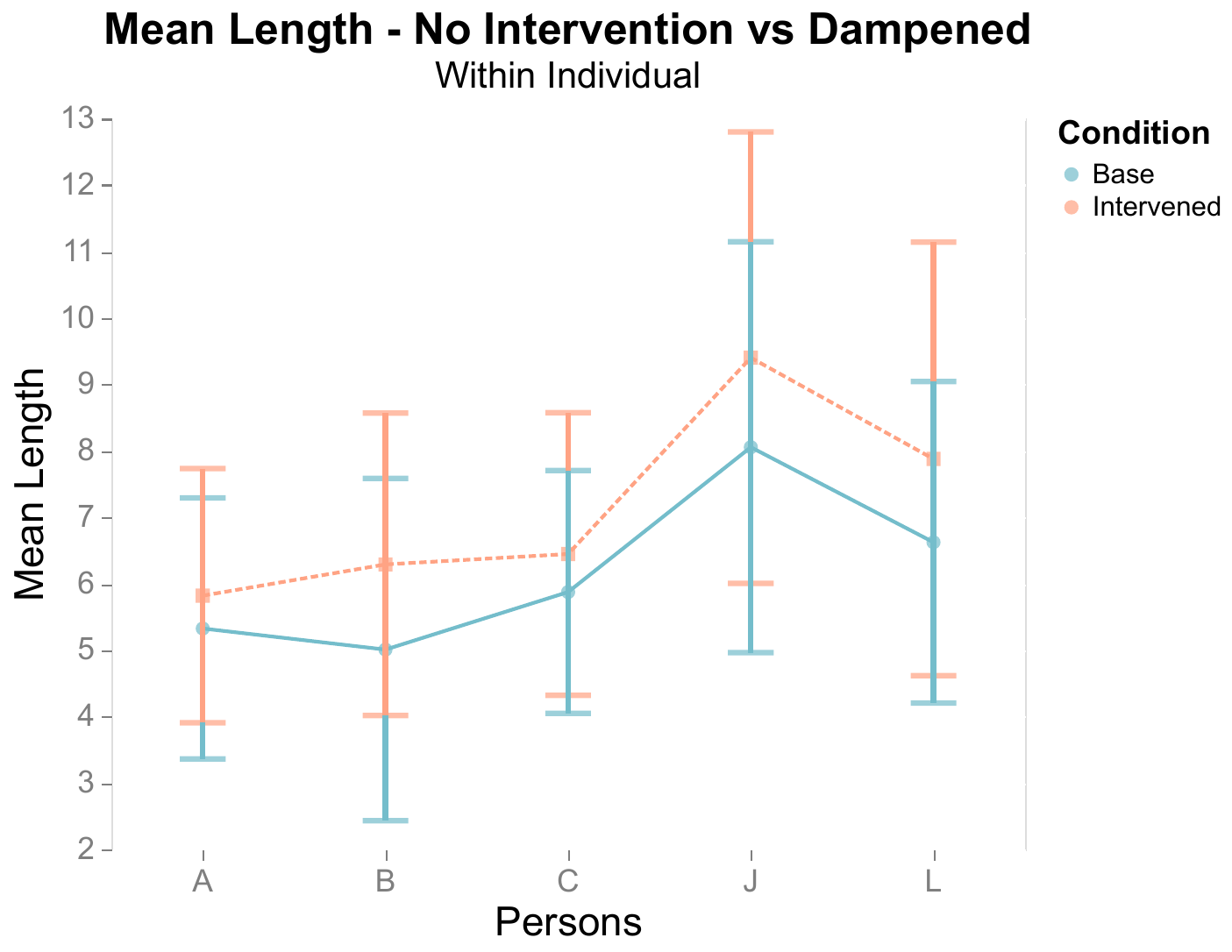}
         \caption{RQA Average Mean Length in an individual.}
         \label{fig:indiv-meanl-err-plt}
     \end{subfigure}
     \hfill
     \caption{Figure shows the RQA Average DET and Mean Length in the gesture signal within each individual. The dampened motion leads to more predictable motion.}
     \label{fig:rqa-both-err-plt}
\end{figure}
\Cref{fig:rqa-both-err-plt} shows the average \%DET per person. The joints on which this analysis was run are: LeftHand, LeftArm, RightHand and RightArm. This analysis was done to capture the synchrony or coordination in the gestures of the same person. The LMEM (summary in \Cref{tab:indiv-det-regression-resultsx10}) shows that the Dampened condition raised \%DET for the Right-Hand by $0.033$ ($p < .001$). The same increase was observed for the Left-Hand. Similarly, MeanLR (\Cref{tab:indiv-meanlr-regression-resultsx10}) for the Right-Hand increased by $2.1$ units in the Dampened condition ($p < .001$). Left-Hand started 0.6 units shorter yet still gained $\approx 1.8$ units under intervention.

\begin{figure}[t]
    \centering
    \includegraphics[width=0.6\linewidth]{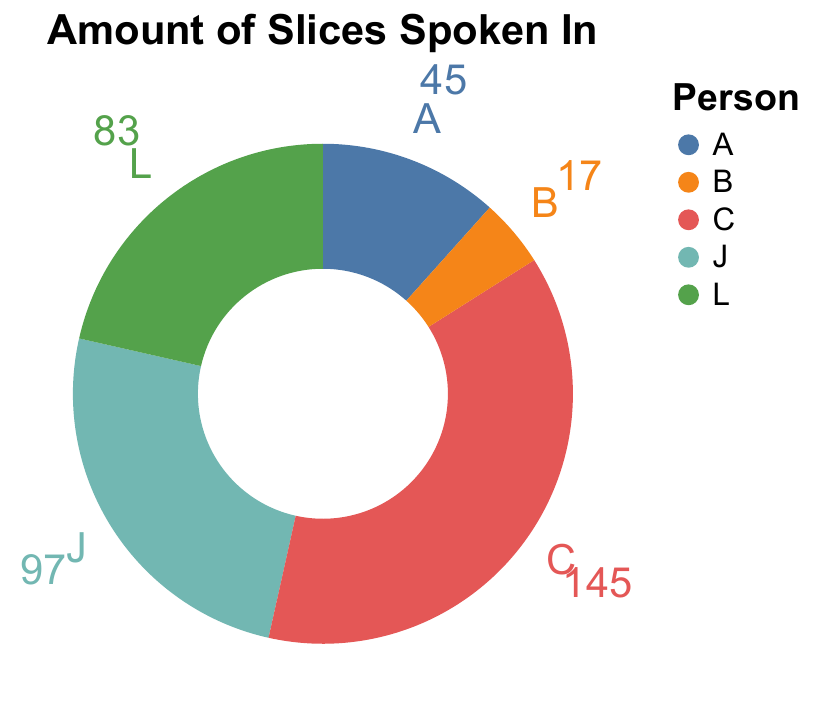}
    \caption{Count of slices a person has spoken in (DnD Session 1).}
    \label{fig:speaking-pie}
\end{figure}

\section{Intra-person Beat Consistency}
\begin{figure}[t]
     \centering
     \begin{subfigure}[t]{0.48\linewidth}
         \centering
         \includegraphics[width=\textwidth]{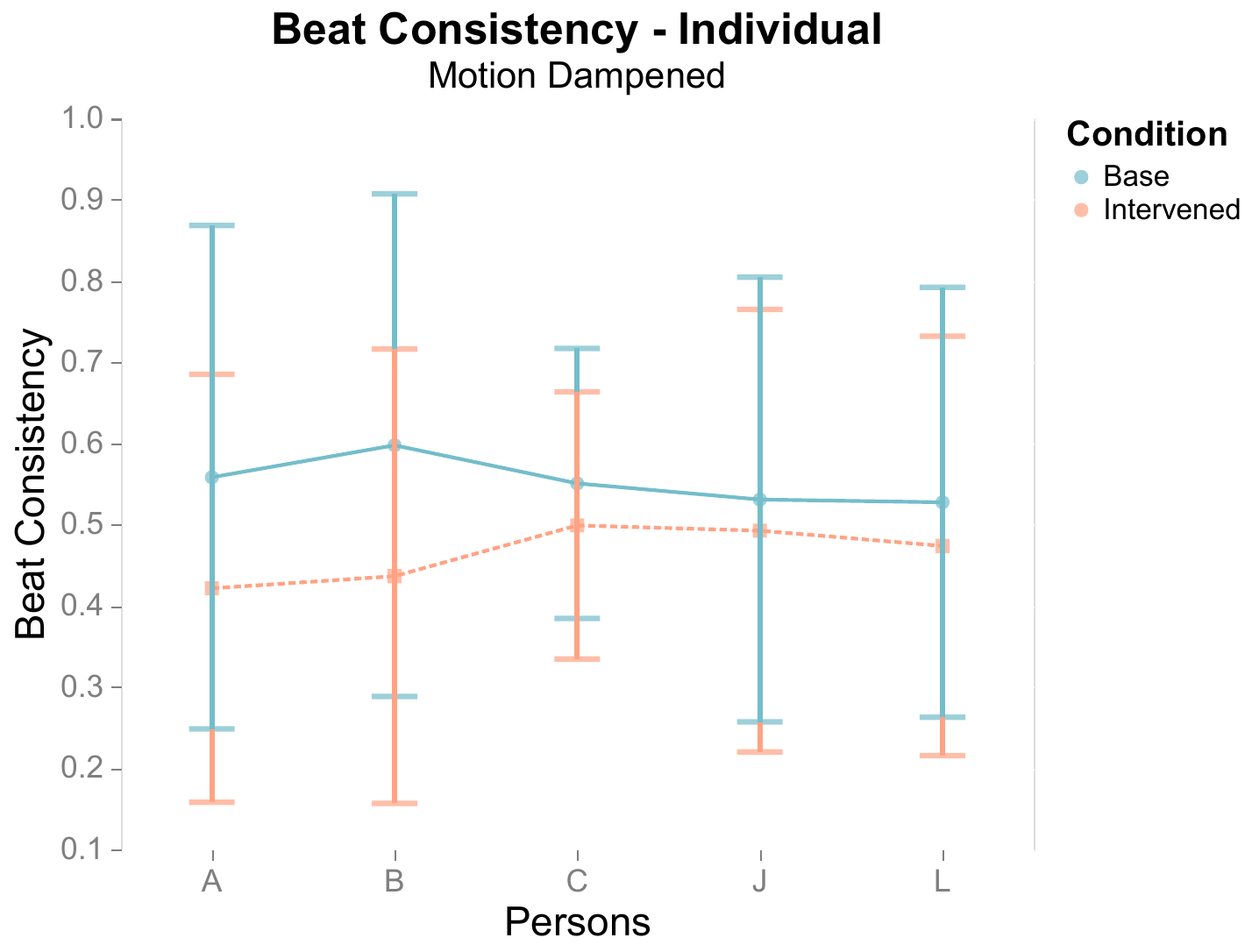}
         \caption{Beat Consistency of an individual with dampened gestures.}
         \label{fig:ind-bc-damp-err-plt}
     \end{subfigure}
     \hfill
     \begin{subfigure}[t]{0.48\linewidth}
         \centering
         \includegraphics[width=\textwidth]{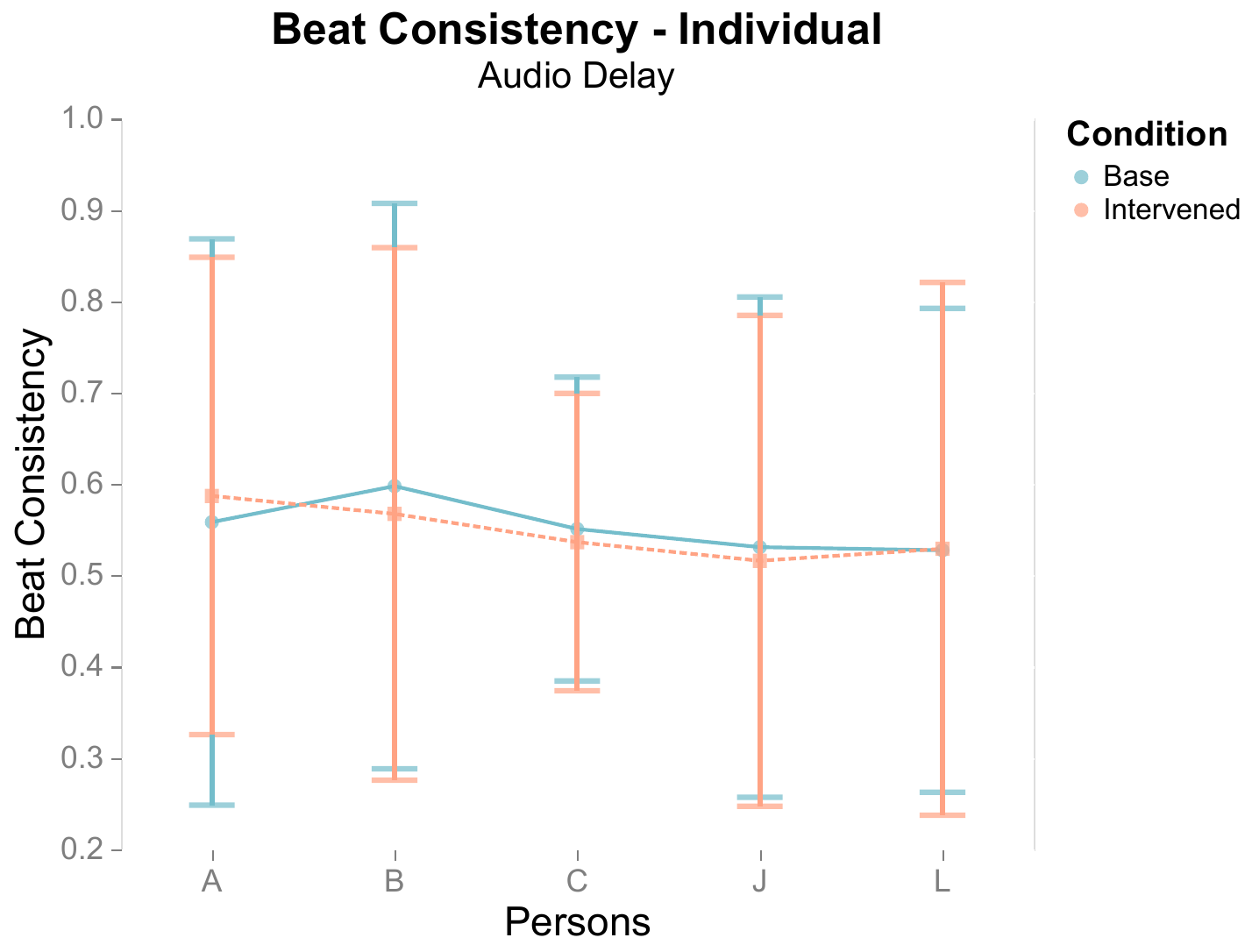}
         \caption{Beat Consistency of an individual with a gesture onset delay.}
         \label{fig:ind-bc-delay-err-plt}
     \end{subfigure}
     \hfill
     \caption{Figure shows the cross-modal self beat consistency of each individual against the manipulated data.}
     \label{fig:ind-bc-plots}
\end{figure}

\Cref{fig:ind-bc-plots} shows the Beat Consistency scores of each individual. Although the LMEM results in the main paper show that the manipulations effect is not strong enough, the error plots do show a slight difference in the means.

\section{Intra-person Pitch Variance SDTW}

\begin{figure}[t]
     \centering
     \begin{subfigure}[t]{0.48\linewidth}
         \centering
         \includegraphics[width=\textwidth]{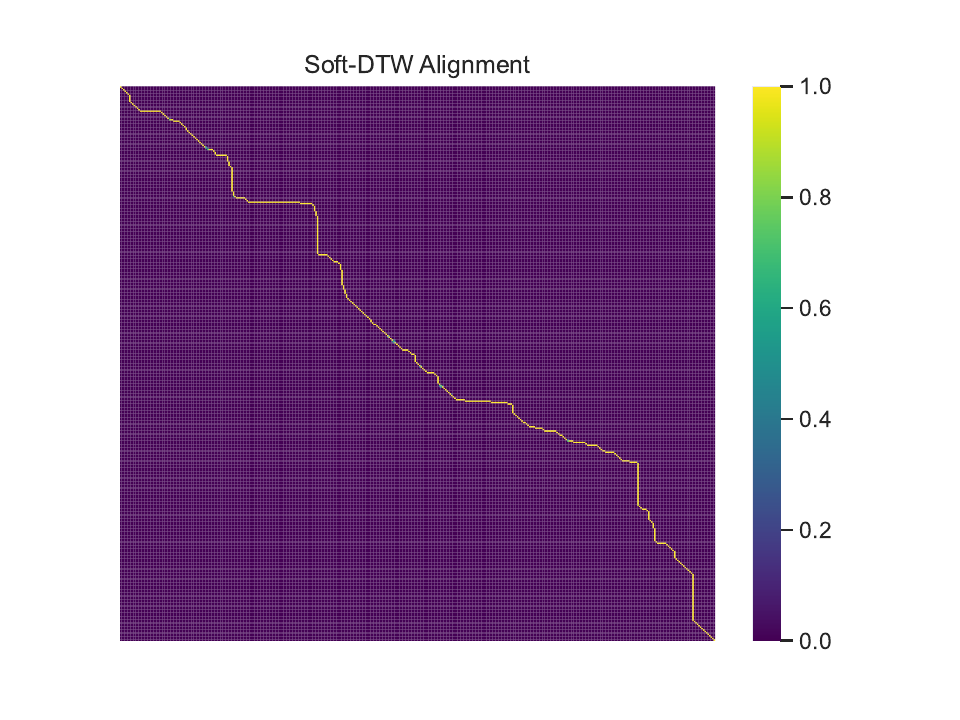}
         \caption{Soft-DTW Alignment Path between two F0 contours, one original and one manipulated.}
         \label{fig:sdtw-alignment-path}
     \end{subfigure}
     \hfill
     \begin{subfigure}[t]{0.48\linewidth}
         \centering
         \includegraphics[width=\textwidth]{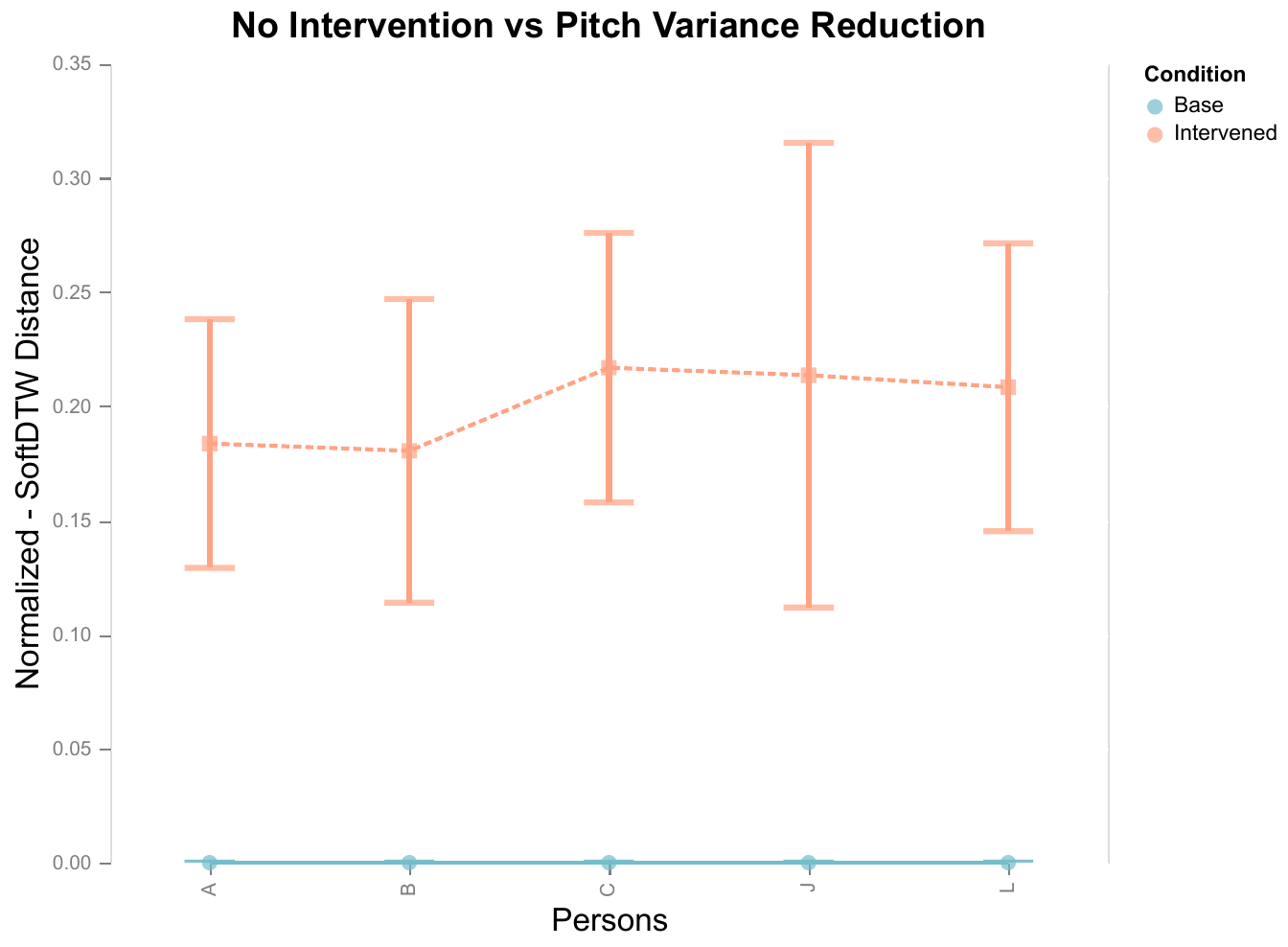}
         \caption{Soft-DTW distances of F0 streams per-person compared against the unaltered F0.}
         \label{fig:ind-pitch-var-err-plt}
     \end{subfigure}
     \hfill
     \caption{Figure shows the cross-modal self beat consistency of each individual against the manipulated data.}
     \label{fig:pitch-var-plots}
\end{figure}

\Cref{fig:sdtw-alignment-path} shows an instance of an alignment matrix between two F0 contours. It calculated between an original F0 contour and an intervened one. For a perfect alignment, the diagonal should have as little deviations as possible which is not the case when using an intervened F0 contour. However, applying SDTW on the original unaltered F0 contour yields a distance of zero as is evidenced in \Cref{fig:ind-pitch-var-err-plt}

\section{Qualtrics Setup}

\begin{figure}[t]
    \centering
    \includegraphics[width=\linewidth]{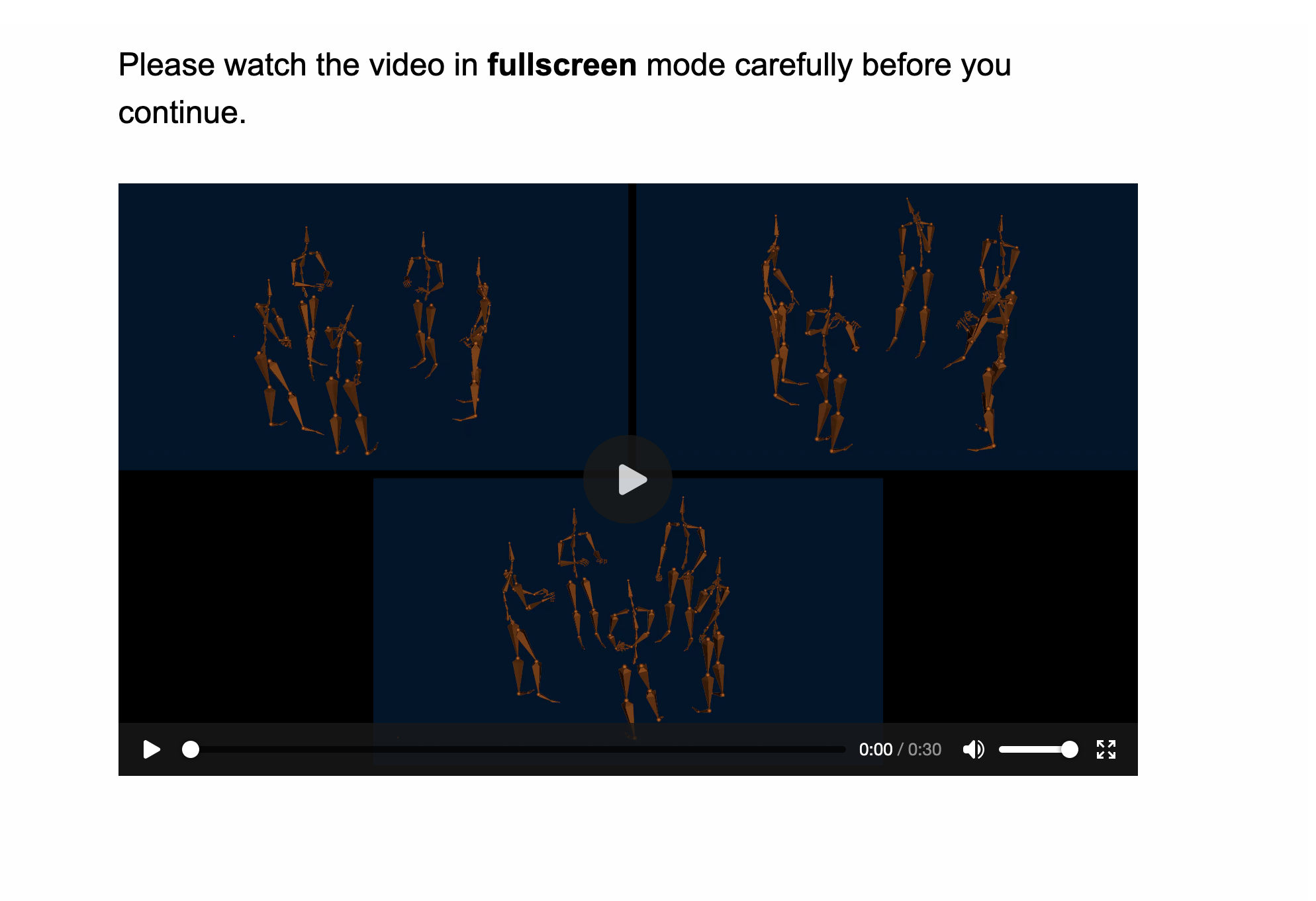}
    \caption{Interface with the video shown to users.}
    \label{fig:qualtrics-group-behav}
\end{figure}

\begin{figure}[t]
    \centering
    \includegraphics[width=\linewidth]{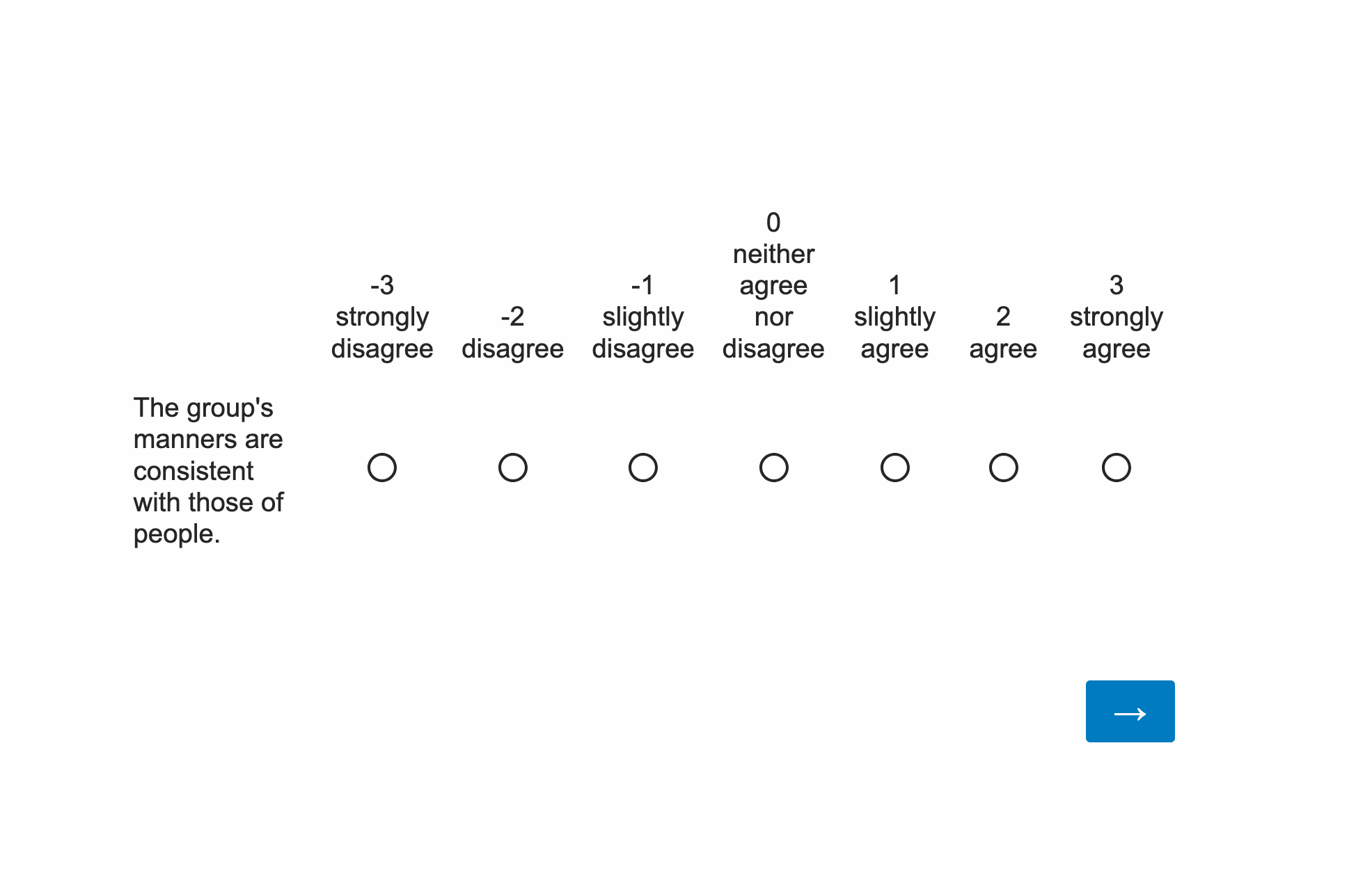}
    \caption{Interface with the questions shown to users.}
    \label{fig:qualtrics-sample-q}
\end{figure}

For our perception study, we used Qualtrics\footnote{\url{https://www.qualtrics.com/}} platform. We setup the ASAQ~\citep{Fitrianie2022-bj} and PCQ~\citep{Raman2023-du}. Since ASAQ is geared towards human-robot interaction (HRI), we slightly modify the question so that they hold semantic relevance in our case (e.g. HRI centric questions to refer to humans instead as the agents). \Cref{fig:qualtrics-group-behav} shows the a thin slice of the group interaction from multiple angles. Before we ask for the ratings the subject is shown a thin slice which they can replay as many times as they like after which they're asked to rate as seen in \Cref{fig:qualtrics-sample-q}. Before the users start rating they're given an instruction:


\setlength{\fboxsep}{10pt}
\setlength{\fboxrule}{2pt}
\fcolorbox{gray!60}{gray!20}{%
    \parbox{0.9\columnwidth}{%
        \textbf{Sample instruction shown to the participants of the user study:}
        
        Please use the set of questions below to indicate your perception of the extent that the group's behaviors appear to be that of real humans, as seen in the video. Each interaction aspect in the below questionnaire should be rated using a seven-point likert scale (strongly disagree (-3) to strongly agree (3)). Read the questions carefully before answering.
        }
    }

\begin{table}[!htbp]
\caption{LMEM for predicting RQA DET (Within Individual). Dampening strength at $10$.}
\label{tab:indiv-det-regression-resultsx10}
\small
\centering
{\tabcolsep=2pt\def\arraystretch{1.0}
\begin{tabularx}{\linewidth}{@{}>{\arraybackslash}l *5{Y}}
\toprule
\textbf{Predictor} & \textbf{Coef.} & \textbf{p-value} & \multicolumn{2}{c}{\(\mathbf{95\%}\) \textbf{CI}} \\
\midrule
Intercept                                           &  0.929 & 0.000 & 0.924 & 0.933 \\
Condition: Intervened                               &  0.026 & 0.000 & 0.021 & 0.030 \\
Joint (vs Right Hand): Left Arm                     &  0.030 & 0.000 & 0.025 & 0.035 \\
Joint (vs Right Hand): Left Hand                    & -0.014 & 0.000 & -0.018 & -0.009 \\
Joint (vs Right Hand): Right Arm                    &  0.030 & 0.000 & 0.025 & 0.035 \\
Intervened × Left Arm                               & -0.026 & 0.000 & -0.032 & -0.019 \\
Intervened × Left Hand                              &  0.001 & 0.882 & -0.006 & 0.007 \\
Intervened × Right Arm                              & -0.026 & 0.000 & -0.032 & -0.019 \\
Group Variance                                      &  0.000 &       &        &       \\
\bottomrule
\end{tabularx}}

\end{table}

\begin{table}[!htbp]
\caption{LMEM for predicting RQA DET (Within Individual). Dampening strength at $20$.}
\label{tab:indiv-det-regression-resultsx20}
\small
\centering
{\tabcolsep=2pt\def\arraystretch{1.0}
\begin{tabularx}{\linewidth}{@{}>{\arraybackslash}l *5{Y}}
\toprule
\textbf{Predictor} & \textbf{Coef.} & \textbf{p-value} & \multicolumn{2}{c}{\(\mathbf{95\%}\) \textbf{CI}} \\
\midrule
Intercept                                           & 0.929 & 0.000 & 0.925 & 0.933 \\
Condition: Intervened                               & 0.033 & 0.000 & 0.028 & 0.038 \\
Joint (vs Right Hand): Left Arm                     & 0.030 & 0.000 & 0.026 & 0.035 \\
Joint (vs Right Hand): Left Hand                    & -0.014 & 0.000 & -0.018 & -0.009 \\
Joint (vs Right Hand): Right Arm                    & 0.030 & 0.000 & 0.025 & 0.035 \\
Intervened × Left Arm                               & -0.033 & 0.000 & -0.040 & -0.027 \\
Intervened × Left Hand                              & 0.003 & 0.386 & -0.004 & 0.009 \\
Intervened × Right Arm                              & -0.033 & 0.000 & -0.040 & -0.027 \\
Group Variance                                      & 0.000 &       &        &       \\
\bottomrule
\end{tabularx}}

\end{table}

\begin{table}[!htbp]
\caption{LMEM for predicting RQA DET (Within Individual). Dampening strength at $30$.}
\label{tab:indiv-det-regression-resultsx30}
\small
\centering
{\tabcolsep=2pt\def\arraystretch{1.0}
\begin{tabularx}{\linewidth}{@{}>{\arraybackslash}l *5{Y}}
\toprule
\textbf{Predictor} & \textbf{Coef.} & \textbf{p-value} & \multicolumn{2}{c}{\(\mathbf{95\%}\) \textbf{CI}} \\
\midrule
Intercept                                           & 0.929 & 0.000 & 0.925 & 0.933 \\
Condition: Intervened                               & 0.036 & 0.000 & 0.032 & 0.041 \\
Joint (vs Right Hand): Left Arm                     & 0.030 & 0.000 & 0.026 & 0.035 \\
Joint (vs Right Hand): Left Hand                    & -0.014 & 0.000 & -0.018 & -0.009 \\
Joint (vs Right Hand): Right Arm                    & 0.030 & 0.000 & 0.025 & 0.035 \\
Intervened × Left Arm                               & -0.036 & 0.000 & -0.043 & -0.030 \\
Intervened × Left Hand                              & 0.004 & 0.194 & -0.002 & 0.011 \\
Intervened × Right Arm                              & -0.036 & 0.000 & -0.043 & -0.030 \\
Group Variance                                      & 0.000 &       &        &       \\
\bottomrule
\end{tabularx}}

\end{table}

\begin{table}[!htbp]
\caption{LMEM for predicting RQA DET (Within Individual). Dampening strength at $40$.}
\label{tab:indiv-det-regression-resultsx40}
\small
\centering
{\tabcolsep=2pt\def\arraystretch{1.0}
\begin{tabularx}{\linewidth}{@{}>{\arraybackslash}l *5{Y}}
\toprule
\textbf{Predictor} & \textbf{Coef.} & \textbf{p-value} & \multicolumn{2}{c}{\(\mathbf{95\%}\) \textbf{CI}} \\
\midrule
Intercept                                           & 0.929 & 0.000 & 0.925 & 0.933 \\
Condition: Intervened                               & 0.038 & 0.000 & 0.034 & 0.043 \\
Joint (vs Right Hand): Left Arm                     & 0.030 & 0.000 & 0.026 & 0.035 \\
Joint (vs Right Hand): Left Hand                    & -0.014 & 0.000 & -0.018 & -0.009 \\
Joint (vs Right Hand): Right Arm                    & 0.030 & 0.000 & 0.025 & 0.034 \\
Intervened × Left Arm                               & -0.038 & 0.000 & -0.044 & -0.032 \\
Intervened × Left Hand                              & 0.005 & 0.140 & -0.002 & 0.011 \\
Intervened × Right Arm                              & -0.038 & 0.000 & -0.044 & -0.032 \\
Group Variance                                      & 0.000 &       &        &       \\
\bottomrule
\end{tabularx}}

\end{table}

\begin{table}[!htbp]
\caption{LMEM for predicting RQA DET (Within Individual). Dampening strength at $50$.}
\label{tab:indiv-det-regression-resultsx50}
\small
\centering
{\tabcolsep=2pt\def\arraystretch{1.0}
\begin{tabularx}{\linewidth}{@{}>{\arraybackslash}l *5{Y}}
\toprule
\textbf{Predictor} & \textbf{Coef.} & \textbf{p-value} & \multicolumn{2}{c}{\(\mathbf{95\%}\) \textbf{CI}} \\
\midrule
Intercept                                           & 0.929 & 0.000 & 0.925 & 0.933 \\
Condition: Intervened                               & 0.039 & 0.000 & 0.035 & 0.044 \\
Joint (vs Right Hand): Left Arm                     & 0.030 & 0.000 & 0.026 & 0.035 \\
Joint (vs Right Hand): Left Hand                    & -0.014 & 0.000 & -0.018 & -0.009 \\
Joint (vs Right Hand): Right Arm                    & 0.030 & 0.000 & 0.025 & 0.034 \\
Intervened × Left Arm                               & -0.039 & 0.000 & -0.046 & -0.033 \\
Intervened × Left Hand                              & 0.005 & 0.099 & -0.001 & 0.012 \\
Intervened × Right Arm                              & -0.039 & 0.000 & -0.046 & -0.033 \\
Group Variance                                      & 0.000 &       &        &       \\
\bottomrule
\end{tabularx}}

\end{table}

\begin{table}[!htbp]
\caption{LMEM for predicting RQA Mean Line Length (Within Individual). Dampening strength at $10$.}
\label{tab:indiv-meanlr-regression-resultsx10}
\small
\centering
{\tabcolsep=2pt\def\arraystretch{1.0}
\begin{tabularx}{\linewidth}{@{}>{\arraybackslash}l *5{Y}}
\toprule
\textbf{Predictor} & \textbf{Coef.} & \textbf{p-value} & \multicolumn{2}{c}{\(\mathbf{95\%}\) \textbf{CI}} \\
\midrule
Intercept                                           & 5.717 & 0.000 & 5.479 & 5.955 \\
Condition: Intervened                               & 1.564 & 0.000 & 1.304 & 1.825 \\
Joint (vs Right Hand): Left Arm                     & 1.179 & 0.000 & 0.918 & 1.439 \\
Joint (vs Right Hand): Left Hand                    & -0.600 & 0.000 & -0.860 & -0.339 \\
Joint (vs Right Hand): Right Arm                    & 1.256 & 0.000 & 0.996 & 1.517 \\
Intervened × Left Arm                               & -1.564 & 0.000 & -1.933 & -1.195 \\
Intervened × Left Hand                              & -0.328 & 0.081 & -0.696 & 0.041 \\
Intervened × Right Arm                              & -1.564 & 0.000 & -1.933 & -1.196 \\
Group Variance                                      & 0.859 &       &        &       \\
\bottomrule
\end{tabularx}}

\end{table}

\begin{table}[!htbp]
\caption{LMEM for predicting RQA Mean Line Length (Within Individual). Dampening strength at $20$.}
\label{tab:indiv-meanlr-regression-resultsx20}
\small
\centering
{\tabcolsep=2pt\def\arraystretch{1.0}
\begin{tabularx}{\linewidth}{@{}>{\arraybackslash}l *5{Y}}
\toprule
\textbf{Predictor} & \textbf{Coef.} & \textbf{p-value} & \multicolumn{2}{c}{\(\mathbf{95\%}\) \textbf{CI}} \\
\midrule
Intercept                                           & 5.717 & 0.000 & 5.473 & 5.961 \\
Condition: Intervened                               & 2.136 & 0.000 & 1.872 & 2.401 \\
Joint (vs Right Hand): Left Arm                     & 1.179 & 0.000 & 0.914 & 1.443 \\
Joint (vs Right Hand): Left Hand                    & -0.600 & 0.000 & -0.864 & -0.335 \\
Joint (vs Right Hand): Right Arm                    & 1.256 & 0.000 & 0.992 & 1.521 \\
Intervened × Left Arm                               & -2.136 & 0.000 & -2.511 & -1.762 \\
Intervened × Left Hand                              & -0.312 & 0.103 & -0.686 & 0.063 \\
Intervened × Right Arm                              & -2.137 & 0.000 & -2.511 & -1.762 \\
Group Variance                                      & 0.928 &       &        &       \\
\bottomrule
\end{tabularx}}

\end{table}

\begin{table}[!htbp]
\caption{LMEM for predicting RQA Mean Line Length (Within Individual). Dampening strength at $30$.}
\label{tab:indiv-meanlr-regression-resultsx30}
\small
\centering
{\tabcolsep=2pt\def\arraystretch{1.0}
\begin{tabularx}{\linewidth}{@{}>{\arraybackslash}l *5{Y}}
\toprule
\textbf{Predictor} & \textbf{Coef.} & \textbf{p-value} & \multicolumn{2}{c}{\(\mathbf{95\%}\) \textbf{CI}} \\
\midrule
Intercept                                           & 5.717 & 0.000 & 5.468 & 5.966 \\
Condition: Intervened                               & 2.553 & 0.000 & 2.283 & 2.822 \\
Joint (vs Right Hand): Left Arm                     & 1.179 & 0.000 & 0.909 & 1.448 \\
Joint (vs Right Hand): Left Hand                    & -0.600 & 0.000 & -0.869 & -0.330 \\
Joint (vs Right Hand): Right Arm                    & 1.256 & 0.000 & 0.987 & 1.526 \\
Intervened × Left Arm                               & -2.552 & 0.000 & -2.933 & -2.171 \\
Intervened × Left Hand                              & -0.315 & 0.105 & -0.696 & 0.066 \\
Intervened × Right Arm                              & -2.553 & 0.000 & -2.934 & -2.172 \\
Group Variance                                      & 0.967 &       &        &       \\
\bottomrule
\end{tabularx}}

\end{table}

\begin{table}[!htbp]
\caption{LMEM for predicting RQA Mean Line Length (Within Individual). Dampening strength at $40$.}
\label{tab:indiv-meanlr-regression-resultsx40}
\small
\centering
{\tabcolsep=2pt\def\arraystretch{1.0}
\begin{tabularx}{\linewidth}{@{}>{\arraybackslash}l *5{Y}}
\toprule
\textbf{Predictor} & \textbf{Coef.} & \textbf{p-value} & \multicolumn{2}{c}{\(\mathbf{95\%}\) \textbf{CI}} \\
\midrule
Intercept                                           & 5.717 & 0.000 & 5.464 & 5.969 \\
Condition: Intervened                               & 2.834 & 0.000 & 2.560 & 3.107 \\
Joint (vs Right Hand): Left Arm                     & 1.179 & 0.000 & 0.905 & 1.452 \\
Joint (vs Right Hand): Left Hand                    & -0.600 & 0.000 & -0.873 & -0.326 \\
Joint (vs Right Hand): Right Arm                    & 1.256 & 0.000 & 0.983 & 1.530 \\
Intervened × Left Arm                               & -2.833 & 0.000 & -3.220 & -2.447 \\
Intervened × Left Hand                              & -0.308 & 0.118 & -0.695 & 0.078 \\
Intervened × Right Arm                              & -2.834 & 0.000 & -3.220 & -2.447 \\
Group Variance                                      & 0.994 &       &        &       \\
\bottomrule
\end{tabularx}}

\end{table}

\begin{table}[!htbp]
\caption{LMEM for predicting RQA Mean Line Length (Within Individual). Dampening strength at $50$.}
\label{tab:indiv-meanlr-regression-resultsx50}
\small
\centering
{\tabcolsep=2pt\def\arraystretch{1.0}
\begin{tabularx}{\linewidth}{@{}>{\arraybackslash}l *5{Y}}
\toprule
\textbf{Predictor} & \textbf{Coef.} & \textbf{p-value} & \multicolumn{2}{c}{\(\mathbf{95\%}\) \textbf{CI}} \\
\midrule
Intercept                                           & 5.737 & 0.000 & 5.482 & 5.992 \\
Condition: Intervened                               & 3.034 & 0.000 & 2.763 & 3.306 \\
Joint (vs Right Hand): Left Arm                     & 1.179 & 0.000 & 0.907 & 1.451 \\
Joint (vs Right Hand): Left Hand                    & -0.600 & 0.000 & -0.872 & -0.327 \\
Joint (vs Right Hand): Right Arm                    & 1.256 & 0.000 & 0.984 & 1.528 \\
Intervened × Left Arm                               & -3.034 & 0.000 & -3.419 & -2.649 \\
Intervened × Left Hand                              & -0.334 & 0.088 & -0.719 & 0.050 \\
Intervened × Right Arm                              & -3.035 & 0.000 & -3.419 & -2.650 \\
Group Variance                                      & 1.062 &       &        &       \\
\bottomrule
\end{tabularx}}

\end{table}

\begin{table}[t]
\caption{LMEM results predicting CRQA \%DET. Dampening strength at \(10\).}
\label{tab:crqa-lmem-det-dampx10-small}
\small
\centering
{\tabcolsep=2pt\def\arraystretch{1.0}
\begin{tabularx}{\linewidth}{@{}>{\arraybackslash}l *5{Y}}
\toprule
\textbf{Predictor} & \textbf{Coef.} & \textbf{p-value} & \multicolumn{2}{c}{\(\mathbf{95\%}\) \textbf{CI}} \\
\midrule
Intercept                                           & -0.093 &     0.000  & -0.134 & -0.052  \\
Condition: Dampened                                   &  0.115 &     0.000  &  0.073 &  0.157  \\
Joint (vs Right Hand): Left Arm                     &  0.180 &     0.000  &  0.137 &  0.222  \\
Joint (vs Right Hand): Right Arm                    &  0.199 &     0.000  &  0.157 &  0.242  \\
Joint (vs Right Hand): Left Hand                    & -0.186 &     0.000  & -0.228 & -0.143  \\
Dampened × Left Arm                                   & -0.115 &     0.000  & -0.175 & -0.055  \\
Dampened × Right Arm                                  & -0.115 &     0.000  & -0.175 & -0.055  \\
Dampened × Left Hand                                  &  0.126 &     0.000  &  0.066 &  0.186  \\
Group Variance                                      &  0.301 &            &        &         \\
\bottomrule
\end{tabularx}}

\end{table}

\begin{table}[t]
\caption{LMEM results predicting CRQA Determinism, with fixed effects for condition, joint, and their interaction.}
\label{tab:crqa-lmem-det-dampx20}
\small
\centering
{\tabcolsep=2pt\def\arraystretch{1.0}
\begin{tabularx}{\linewidth}{@{}>{\arraybackslash}l *5{Y}}
\toprule
\textbf{Predictor} & \textbf{Coef.} & \textbf{p-value} & \multicolumn{2}{c}{\(\mathbf{95\%}\) \textbf{CI}} \\
\midrule
Intercept                                           & -0.110 &     0.000  & -0.151 & -0.069  \\
Condition: Dampened                                   &  0.175 &     0.000  &  0.132 &  0.218  \\
Joint (vs Right Hand): Left Arm                     &  0.180 &     0.000  &  0.137 &  0.223  \\
Joint (vs Right Hand): Left Hand                    & -0.186 &     0.000  & -0.229 & -0.143  \\
Joint (vs Right Hand): Right Arm                    &  0.200 &     0.000  &  0.157 &  0.243  \\
Dampened × Left Arm                                   & -0.175 &     0.000  & -0.236 & -0.114  \\
Dampened × Left Hand                                  &  0.142 &     0.000  &  0.082 &  0.203  \\
Dampened × Right Arm                                  & -0.175 &     0.000  & -0.236 & -0.114  \\
Group Variance                                      &  0.283 &            &        &         \\
\bottomrule
\end{tabularx}}

\end{table}

\begin{table}[t]
\caption{LMEM results predicting CRQA Determinism with fixed effects for condition, joint, and their interaction.}
\label{tab:crqa-lmem-det-dampx30}
\small
\centering
{\tabcolsep=2pt\def\arraystretch{1.0}
\begin{tabularx}{\linewidth}{@{}>{\arraybackslash}l *5{Y}}
\toprule
\textbf{Predictor} & \textbf{Coef.} & \textbf{p-value} & \multicolumn{2}{c}{\(\mathbf{95\%}\) \textbf{CI}} \\
\midrule
Intercept                                           & -0.111 &     0.000  & -0.151 & -0.070  \\
Condition: Dampened                                   &  0.175 &     0.000  &  0.132 &  0.218  \\
Joint (vs Right Hand): Left Hand                    & -0.183 &     0.000  & -0.226 & -0.140  \\
Joint (vs Right Hand): Left Arm                     &  0.177 &     0.000  &  0.134 &  0.220  \\
Joint (vs Right Hand): Right Arm                    &  0.196 &     0.000  &  0.153 &  0.240  \\
Dampened × Left Hand                                  &  0.154 &     0.000  &  0.093 &  0.215  \\
Dampened × Left Arm                                   & -0.175 &     0.000  & -0.236 & -0.114  \\
Dampened × Right Arm                                  & -0.175 &     0.000  & -0.236 & -0.114  \\
Group Variance                                      &  0.274 &            &        &         \\
\bottomrule
\end{tabularx}}

\end{table}

\begin{table}[t]
\caption{LMEM results predicting CRQA Determinism with fixed effects for condition, joint, and their interaction.}
\label{tab:crqa-lmem-det-dampx40}
\small
\centering
{\tabcolsep=2pt\def\arraystretch{1.0}
\begin{tabularx}{\linewidth}{@{}>{\arraybackslash}l *5{Y}}
\toprule
\textbf{Predictor} & \textbf{Coef.} & \textbf{p-value} & \multicolumn{2}{c}{\(\mathbf{95\%}\) \textbf{CI}} \\
\midrule
Intercept                                           & -0.118 &     0.000  & -0.159 & -0.077  \\
Condition: Dampened                                   &  0.220 &     0.000  &  0.177 &  0.264  \\
Joint (vs Right Hand): Right Arm                    &  0.199 &     0.000  &  0.156 &  0.242  \\
Joint (vs Right Hand): Left Arm                     &  0.180 &     0.000  &  0.137 &  0.223  \\
Joint (vs Right Hand): Left Hand                    & -0.186 &     0.000  & -0.229 & -0.143  \\
Dampened × Right Arm                                  & -0.220 &     0.000  & -0.282 & -0.159  \\
Dampened × Left Arm                                   & -0.220 &     0.000  & -0.281 & -0.159  \\
Dampened × Left Hand                                  &  0.117 &     0.000  &  0.056 &  0.178  \\
Group Variance                                      &  0.273 &            &        &         \\
\bottomrule
\end{tabularx}}

\end{table}

\begin{table}[t]
\caption{LMEM results predicting CRQA Determinism with fixed effects for condition, joint, and their interaction.}
\label{tab:crqa-lmem-det-dampx50}
\small
\centering
{\tabcolsep=2pt\def\arraystretch{1.0}
\begin{tabularx}{\linewidth}{@{}>{\arraybackslash}l *5{Y}}
\toprule
\textbf{Predictor} & \textbf{Coef.} & \textbf{p-value} & \multicolumn{2}{c}{\(\mathbf{95\%}\) \textbf{CI}} \\
\midrule
Intercept                                           & -0.201 &     0.000  & -0.241 & -0.161  \\
Condition: Dampened                                   &  0.460 &     0.000  &  0.417 &  0.503  \\
Joint (vs Right Hand): Left Arm                     &  0.244 &     0.000  &  0.201 &  0.287  \\
Joint (vs Right Hand): Right Arm                    &  0.273 &     0.000  &  0.230 &  0.316  \\
Joint (vs Right Hand): Left Hand                    & -0.258 &     0.000  & -0.302 & -0.215  \\
Dampened × Left Arm                                   & -0.460 &     0.000  & -0.521 & -0.399  \\
Dampened × Right Arm                                  & -0.460 &     0.000  & -0.521 & -0.399  \\
Dampened × Left Hand                                  &  0.170 &     0.000  &  0.109 &  0.231  \\
Group Variance                                      &  0.246 &            &        &         \\
\bottomrule
\end{tabularx}}

\end{table}

\begin{table}[htbp]
\caption{LMEM results predicting CRQA MeanLR. Dampening strength at \(10\).}
\label{tab:crqa-lmem-meanlen-dampx10}
\small
\centering
{\tabcolsep=2pt\def\arraystretch{1.0}
\begin{tabularx}{\linewidth}{@{}>{\arraybackslash}l *5{Y}}
\toprule
\textbf{Predictor} & \textbf{Coef.} & \textbf{p-value} & \multicolumn{2}{c}{\(\mathbf{95\%}\) \textbf{CI}} \\
\midrule
Intercept                                           & -0.169 &     0.000  & -0.210 & -0.128  \\
Condition: Dampened                                   &  0.451 &     0.000  &  0.410 &  0.492  \\
Joint (vs Right Hand): Left Arm                     &  0.204 &     0.000  &  0.163 &  0.245  \\
Joint (vs Right Hand): Left Hand                    & -0.172 &     0.000  & -0.213 & -0.131  \\
Joint (vs Right Hand): Right Arm                    &  0.222 &     0.000  &  0.181 &  0.263  \\
Dampened × Left Arm                                   & -0.451 &     0.000  & -0.509 & -0.393  \\
Dampened × Left Hand                                  & -0.059 &     0.046  & -0.117 & -0.001  \\
Dampened × Right Arm                                  & -0.451 &     0.000  & -0.509 & -0.393  \\
Group Variance                                      &  0.334 &            &        &         \\
\bottomrule
\end{tabularx}}

\end{table}

\begin{table}[htbp]
\caption{LMEM results predicting CRQA Mean Length with fixed effects for condition, joint, and their interaction.}
\label{tab:crqa-lmem-meanlen-dampx20}
\small
\centering
{\tabcolsep=2pt\def\arraystretch{1.0}
\begin{tabularx}{\linewidth}{@{}>{\arraybackslash}l *5{Y}}
\toprule
\textbf{Predictor} & \textbf{Coef.} & \textbf{p-value} & \multicolumn{2}{c}{\(\mathbf{95\%}\) \textbf{CI}} \\
\midrule
Intercept                                           & -0.210 &     0.000  & -0.251 & -0.170  \\
Condition: Dampened                                   &  0.646 &     0.000  &  0.605 &  0.688  \\
Joint (vs Right Hand): Left Arm                     &  0.197 &     0.000  &  0.156 &  0.238  \\
Joint (vs Right Hand): Right Arm                    &  0.214 &     0.000  &  0.173 &  0.255  \\
Joint (vs Right Hand): Left Hand                    & -0.166 &     0.000  & -0.207 & -0.125  \\
Dampened × Left Arm                                   & -0.646 &     0.000  & -0.704 & -0.588  \\
Dampened × Right Arm                                  & -0.646 &     0.000  & -0.704 & -0.589  \\
Dampened × Left Hand                                  & -0.100 &     0.001  & -0.158 & -0.042  \\
Group Variance                                      &  0.311 &            &        &         \\
\bottomrule
\end{tabularx}}

\end{table}

\begin{table}[htbp]
\caption{LMEM results predicting CRQA MeanLR. Dampening strength at \(30\).}
\label{tab:crqa-lmem-meanlen-dampx30}
\small
\centering
{\tabcolsep=2pt\def\arraystretch{1.0}
\begin{tabularx}{\linewidth}{@{}>{\arraybackslash}l *5{Y}}
\toprule
\textbf{Predictor} & \textbf{Coef.} & \textbf{p-value} & \multicolumn{2}{c}{\(\mathbf{95\%}\) \textbf{CI}} \\
\midrule
Intercept                                           & -0.233 &     0.000  & -0.273 & -0.193  \\
Condition: Dampened                                   &  0.754 &     0.000  &  0.713 &  0.795  \\
Joint (vs Right Hand): Left Hand                    & -0.160 &     0.000  & -0.201 & -0.119  \\
Joint (vs Right Hand): Right Arm                    &  0.207 &     0.000  &  0.166 &  0.248  \\
Joint (vs Right Hand): Left Arm                     &  0.190 &     0.000  &  0.149 &  0.231  \\
Dampened × Left Hand                                  & -0.117 &     0.000  & -0.175 & -0.059  \\
Dampened × Right Arm                                  & -0.754 &     0.000  & -0.812 & -0.696  \\
Dampened × Left Arm                                   & -0.754 &     0.000  & -0.812 & -0.696  \\
Group Variance                                      &  0.296 &            &        &         \\
\bottomrule
\end{tabularx}}

\end{table}

\begin{table}[htbp]
\caption{LMEM results predicting CRQA MeanLR. Dampening strength at \(40\).}
\label{tab:crqa-lmem-meanlen-dampx40}
\small
\centering
{\tabcolsep=2pt\def\arraystretch{1.0}
\begin{tabularx}{\linewidth}{@{}>{\arraybackslash}l *5{Y}}
\toprule
\textbf{Predictor} & \textbf{Coef.} & \textbf{p-value} & \multicolumn{2}{c}{\(\mathbf{95\%}\) \textbf{CI}} \\
\midrule
Intercept                                           & -0.246 &     0.000  & -0.285 & -0.206  \\
Condition: Dampened                                   &  0.809 &     0.000  &  0.768 &  0.849  \\
Joint (vs Right Hand): Right Arm                    &  0.202 &     0.000  &  0.161 &  0.243  \\
Joint (vs Right Hand): Left Arm                     &  0.185 &     0.000  &  0.145 &  0.226  \\
Joint (vs Right Hand): Left Hand                    & -0.156 &     0.000  & -0.197 & -0.115  \\
Dampened × Right Arm                                  & -0.809 &     0.000  & -0.866 & -0.751  \\
Dampened × Left Arm                                   & -0.809 &     0.000  & -0.866 & -0.751  \\
Dampened × Left Hand                                  & -0.115 &     0.000  & -0.173 & -0.057  \\
Group Variance                                      &  0.290 &            &        &         \\
\bottomrule
\end{tabularx}}

\end{table}

\begin{table}[htbp]
\caption{LMEM results predicting CRQA MeanLR. Dampening strength at \(50\).}
\label{tab:crqa-lmem-meanlen-dampx50}
\small
\centering
{\tabcolsep=2pt\def\arraystretch{1.0}
\begin{tabularx}{\linewidth}{@{}>{\arraybackslash}l *5{Y}}
\toprule
\textbf{Predictor} & \textbf{Coef.} & \textbf{p-value} & \multicolumn{2}{c}{\(\mathbf{95\%}\) \textbf{CI}} \\
\midrule
Intercept                                           & -0.238 &     0.000  & -0.277 & -0.198  \\
Condition: Dampened                                   &  0.853 &     0.000  &  0.814 &  0.893  \\
Joint (vs Right Hand): Left Arm                     &  0.160 &     0.000  &  0.121 &  0.200  \\
Joint (vs Right Hand): Right Arm                    &  0.186 &     0.000  &  0.146 &  0.226  \\
Joint (vs Right Hand): Left Hand                    & -0.212 &     0.000  & -0.251 & -0.172  \\
Dampened × Left Arm                                   & -0.853 &     0.000  & -0.909 & -0.797  \\
Dampened × Right Arm                                  & -0.853 &     0.000  & -0.909 & -0.797  \\
Dampened × Left Hand                                  & -0.075 &     0.009  & -0.132 & -0.019  \\
Group Variance                                      &  0.305 &            &        &         \\
\bottomrule
\end{tabularx}}

\end{table}


\clearpage



\begin{table}[!htbp]
\caption{LMEM predicting intra-person Beat Consistency for the motion dampening manipulation; strength of $10$.}
\label{tab:indiv-bc-dampened-lmemx10}
\small
\centering
{\tabcolsep=4pt\def\arraystretch{1.0}
\begin{tabularx}{\linewidth}{@{}>{\arraybackslash}l *5{Y}}
\toprule
\textbf{Predictor} & \textbf{Coef.} & \textbf{p-value} & \multicolumn{2}{c}{\(\mathbf{95\%}\) \textbf{CI}} \\
\midrule
Intercept             & 0.543 & 0.000 & 0.518 & 0.568 \\
Condition: Intervened & -0.018 & 0.273 & -0.049 & 0.014 \\
Group Variance        & 0.000 &       &        &       \\
\bottomrule
\end{tabularx}}

\end{table}

\begin{table}[htbp]
\caption{LMEM predicting intra-person Beat Consistency for the motion dampening manipulation; strength of 20}
\label{tab:indiv-bc-dampened-lmemx20}
\small
\centering
{\tabcolsep=4pt\def\arraystretch{1.0}
\begin{tabularx}{\linewidth}{@{}>{\arraybackslash}l *5{Y}}
\toprule
\textbf{Predictor} & \textbf{Coef.} & \textbf{p-value} & \multicolumn{2}{c}{\(\mathbf{95\%}\) \textbf{CI}} \\
\midrule
Intercept             & 0.544 & 0.000 & 0.520 & 0.567 \\
Condition: Intervened & -0.064 & 0.000 & -0.095 & -0.032 \\
Group Variance        & 0.000 &       &        &       \\
\bottomrule
\end{tabularx}}

\end{table}

\begin{table}[htbp]
\caption{LMEM predicting intra-person Beat Consistency for the motion dampening manipulation; strength of 30}
\label{tab:indiv-bc-dampened-lmemx30}
\small
\centering
{\tabcolsep=4pt\def\arraystretch{1.0}
\begin{tabularx}{\linewidth}{@{}>{\arraybackslash}l *5{Y}}
\toprule
\textbf{Predictor} & \textbf{Coef.} & \textbf{p-value} & \multicolumn{2}{c}{\(\mathbf{95\%}\) \textbf{CI}} \\
\midrule
Intercept             & 0.544 & 0.000 & 0.521 & 0.566 \\
Condition: Intervened & -0.098 & 0.000 & -0.129 & -0.066 \\
Group Variance        & 0.000 &       &        &       \\
\bottomrule
\end{tabularx}}

\end{table}

\begin{table}[htbp]
\caption{LMEM predicting intra-person Beat Consistency for the motion dampening manipulation; strength of 40}
\label{tab:indiv-bc-dampened-lmemx40}
\small
\centering
{\tabcolsep=4pt\def\arraystretch{1.0}
\begin{tabularx}{\linewidth}{@{}>{\arraybackslash}l *5{Y}}
\toprule
\textbf{Predictor} & \textbf{Coef.} & \textbf{p-value} & \multicolumn{2}{c}{\(\mathbf{95\%}\) \textbf{CI}} \\
\midrule
Intercept             & 0.543 & 0.000 & 0.519 & 0.567 \\
Condition: Intervened & -0.142 & 0.000 & -0.174 & -0.111 \\
Group Variance        & 0.000 &       &        &       \\
\bottomrule
\end{tabularx}}

\end{table}

\begin{table}[htbp]
\caption{LMEM predicting intra-person Beat Consistency for the motion dampening manipulation; strength of 50.}
\label{tab:indiv-bc-dampened-lmemx50}
\small
\centering
{\tabcolsep=4pt\def\arraystretch{1.0}
\begin{tabularx}{\linewidth}{@{}>{\arraybackslash}l *5{Y}}
\toprule
\textbf{Predictor} & \textbf{Coef.} & \textbf{p-value} & \multicolumn{2}{c}{\(\mathbf{95\%}\) \textbf{CI}} \\
\midrule
Intercept             & 0.544 & 0.000 & 0.521 & 0.567 \\
Condition: Intervened & -0.196 & 0.000 & -0.228 & -0.165 \\
Group Variance        & 0.000 &       &        &       \\
\bottomrule
\end{tabularx}}

\end{table}

\begin{table}[htbp]
\caption{LMEM predicting intra-person Beat Consistency for the audio delay manipulation; delay set at \(0.15\)s.}
\label{tab:indiv-bc-lmem-audiox0.15}
\small
\centering
{\tabcolsep=4pt\def\arraystretch{1.0}
\begin{tabularx}{\linewidth}{@{}>{\arraybackslash}l *5{Y}}
\toprule
\textbf{Predictor} & \textbf{Coef.} & \textbf{p-value} & \multicolumn{2}{c}{\(\mathbf{95\%}\) \textbf{CI}} \\
\midrule
Intercept             & 0.544 & 0.000 & 0.521 & 0.566 \\
Condition: Intervened & 0.004 & 0.780 & -0.027 & 0.036 \\
Group Variance        & 0.000 &       &        &       \\
\bottomrule
\end{tabularx}}

\end{table}

\begin{table}[htbp]
\caption{LMEM predicting intra-person Beat Consistency for the audio delay manipulation; delay set at \(0.25\)s.}
\label{tab:indiv-bc-lmem-audiox0.25}
\small
\centering
{\tabcolsep=4pt\def\arraystretch{1.0}
\begin{tabularx}{\linewidth}{@{}>{\arraybackslash}l *5{Y}}
\toprule
\textbf{Predictor} & \textbf{Coef.} & \textbf{p-value} & \multicolumn{2}{c}{\(\mathbf{95\%}\) \textbf{CI}} \\
\midrule
Intercept             & 0.544 & 0.000 & 0.508 & 0.579 \\
Condition: Intervened & 0.009 & 0.580 & -0.022 & 0.040 \\
Group Variance        & 0.000 &       &        &       \\
\bottomrule
\end{tabularx}}

\end{table}

\begin{table}[htbp]
\caption{LMEM predicting intra-person Beat Consistency for the audio delay manipulation; delay set at \(0.50\)s.}
\label{tab:indiv-bc-lmem-audiox0.50}
\small
\centering
{\tabcolsep=4pt\def\arraystretch{1.0}
\begin{tabularx}{\linewidth}{@{}>{\arraybackslash}l *5{Y}}
\toprule
\textbf{Predictor} & \textbf{Coef.} & \textbf{p-value} & \multicolumn{2}{c}{\(\mathbf{95\%}\) \textbf{CI}} \\
\midrule
Intercept             & 0.544 & 0.000 & 0.520 & 0.567 \\
Condition: Intervened & -0.009 & 0.561 & -0.041 & 0.022 \\
Group Variance        & 0.000 &       &        &       \\
\bottomrule
\end{tabularx}}

\end{table}

\begin{table}[htbp]
\caption{LMEM predicting intra-person Beat Consistency for the audio delay manipulation; delay set at \(0.75\)s.}
\label{tab:indiv-bc-lmem-audiox0.75}
\small
\centering
{\tabcolsep=4pt\def\arraystretch{1.0}
\begin{tabularx}{\linewidth}{@{}>{\arraybackslash}l *5{Y}}
\toprule
\textbf{Predictor} & \textbf{Coef.} & \textbf{p-value} & \multicolumn{2}{c}{\(\mathbf{95\%}\) \textbf{CI}} \\
\midrule
Intercept             & 0.544 & 0.000 & 0.517 & 0.570 \\
Condition: Intervened & -0.005 & 0.756 & -0.036 & 0.026 \\
Group Variance        & 0.000 &       &        &       \\
\bottomrule
\end{tabularx}}

\end{table}

\begin{table}[htbp]
\caption{LMEM predicting intra-person Beat Consistency for the audio delay manipulation; delay set at \(1.40\)s.}
\label{tab:indiv-bc-lmem-audiox1.40}
\small
\centering
{\tabcolsep=4pt\def\arraystretch{1.0}
\begin{tabularx}{\linewidth}{@{}>{\arraybackslash}l *5{Y}}
\toprule
\textbf{Predictor} & \textbf{Coef.} & \textbf{p-value} & \multicolumn{2}{c}{\(\mathbf{95\%}\) \textbf{CI}} \\
\midrule
Intercept             & 0.544 & 0.000 & 0.521 & 0.567 \\
Condition: Intervened & -0.007 & 0.671 & -0.038 & 0.024 \\
Group Variance        & 0.000 &       &        &       \\
\bottomrule
\end{tabularx}}

\end{table}

\clearpage

\begin{table}[t]
\caption{LMEM predicting inter-person Beat Consistency for motion dampening strength of $10$.}
\label{tab:cross-bc-dampened-lmemx10}
\small
\centering
{\tabcolsep=2pt\def\arraystretch{1.0}
\begin{tabularx}{\linewidth}{@{}>{\arraybackslash}l *5{Y}}
\toprule
\textbf{Predictor} & \textbf{Coef.} & \textbf{p-value} & \multicolumn{2}{c}{\(\mathbf{95\%}\) \textbf{CI}} \\
\midrule
Intercept                   & 0.559 & 0.000 & 0.546 & 0.571 \\
Condition: Intervened       & 0.013 & 0.030 & 0.001 & 0.025 \\
Group Variance              & 0.021 &       &       &       \\
\bottomrule
\end{tabularx}}
\end{table}

\begin{table}[t]
\caption{LMEM predicting inter-person Beat Consistency for motion dampening strength of $20$.}
\label{tab:cross-bc-dampened-lmemx20}
\small
\centering
{\tabcolsep=2pt\def\arraystretch{1.0}
\begin{tabularx}{\linewidth}{@{}>{\arraybackslash}l *5{Y}}
\toprule
\textbf{Predictor} & \textbf{Coef.} & \textbf{p-value} & \multicolumn{2}{c}{\(\mathbf{95\%}\) \textbf{CI}} \\
\midrule
Intercept                   & 0.564 & 0.000 & 0.552 & 0.576 \\
Condition: Intervened       & -0.049 & 0.000 & -0.062 & -0.036 \\
Group Variance              & 0.019 &       &        &        \\
\bottomrule
\end{tabularx}}

\end{table}

\begin{table}[t]
\caption{LMEM predicting inter-person Beat Consistency for motion dampening strength of $30$.}
\label{tab:cross-bc-dampened-lmemx30}
\small
\centering
{\tabcolsep=2pt\def\arraystretch{1.0}
\begin{tabularx}{\linewidth}{@{}>{\arraybackslash}l *5{Y}}
\toprule
\textbf{Predictor} & \textbf{Coef.} & \textbf{p-value} & \multicolumn{2}{c}{\(\mathbf{95\%}\) \textbf{CI}} \\
\midrule
Intercept                   & 0.558 & 0.000 & 0.545 & 0.571 \\
Condition: Intervened       & -0.089 & 0.000 & -0.102 & -0.075 \\
Group Variance              & 0.019 &       &        &        \\
\bottomrule
\end{tabularx}}

\end{table}

\begin{table}[t]
\caption{LMEM predicting inter-person Beat Consistency for motion dampening strength of $40$.}
\label{tab:cross-bc-dampened-lmemx40}
\small
\centering
{\tabcolsep=2pt\def\arraystretch{1.0}
\begin{tabularx}{\linewidth}{@{}>{\arraybackslash}l *5{Y}}
\toprule
\textbf{Predictor} & \textbf{Coef.} & \textbf{p-value} & \multicolumn{2}{c}{\(\mathbf{95\%}\) \textbf{CI}} \\
\midrule
Intercept                   & 0.557 & 0.000 & 0.544 & 0.569 \\
Condition: Intervened       & -0.150 & 0.000 & -0.164 & -0.137 \\
Group Variance              & 0.018 &       &        &        \\
\bottomrule
\end{tabularx}}

\end{table}

\begin{table}[t]
\caption{LMEM predicting inter-person Beat Consistency for motion dampening strength of $50$.}
\label{tab:cross-bc-dampened-lmemx50}
\small
\centering
{\tabcolsep=2pt\def\arraystretch{1.0}
\begin{tabularx}{\linewidth}{@{}>{\arraybackslash}l *5{Y}}
\toprule
\textbf{Predictor} & \textbf{Coef.} & \textbf{p-value} & \multicolumn{2}{c}{\(\mathbf{95\%}\) \textbf{CI}} \\
\midrule
Intercept                   & 0.572 & 0.000 & 0.560 & 0.585 \\
Condition: Intervened       & -0.203 & 0.000 & -0.217 & -0.189 \\
Group Variance              & 0.016 &       &        &        \\
\bottomrule
\end{tabularx}}

\end{table}

\begin{table}[htbp]
\caption{LMEM predicting inter-person Beat Consistency for the audio delay manipulation. Delay strength at $0.15$s.}
\label{tab:lmem-cross-bc-audio-delayx0.15}
\small
\centering
{\tabcolsep=4pt\def\arraystretch{1.0}
\begin{tabularx}{\linewidth}{@{}>{\arraybackslash}l *5{Y}}
\toprule
\textbf{Predictor} & \textbf{Coef.} & \textbf{p-value} & \multicolumn{2}{c}{\(\mathbf{95\%}\) \textbf{CI}} \\
\midrule
Intercept             & 0.569 & 0.000 & 0.556 & 0.581 \\
Condition: Intervened & 0.001 & 0.458 & -0.002 & 0.004 \\
Group Variance        & 0.045 &       &        &       \\
\bottomrule
\end{tabularx}}

\end{table}

\begin{table}[htbp]
\caption{LMEM predicting inter-person Beat Consistency for the audio delay manipulation. Delay set to \(0.25\)s.}
\label{tab:lmem-cross-bc-audio-delayx0.25}
\small
\centering
{\tabcolsep=2pt\def\arraystretch{1.0}
\begin{tabularx}{\linewidth}{@{}>{\arraybackslash}l *5{Y}}
\toprule
\textbf{Predictor} & \textbf{Coef.} & \textbf{p-value} & \multicolumn{2}{c}{\(\mathbf{95\%}\) \textbf{CI}} \\
\midrule
Intercept             & 0.565 & 0.000 & 0.552 & 0.578 \\
Condition: Intervened & -0.004 & 0.041 & -0.008 & -0.000 \\
Group Variance        & 0.047 &       &        &       \\
\bottomrule
\end{tabularx}}

\end{table}

\begin{table}[htbp]
\caption{LMEM predicting inter-person Beat Consistency for the audio delay manipulation. Delay set at $0.50$s}
\label{tab:lmem-cross-bc-audio-delayx0.50}
\small
\centering
{\tabcolsep=2pt\def\arraystretch{1.0}
\begin{tabularx}{\linewidth}{@{}>{\arraybackslash}l *5{Y}}
\toprule
\textbf{Predictor} & \textbf{Coef.} & \textbf{p-value} & \multicolumn{2}{c}{\(\mathbf{95\%}\) \textbf{CI}} \\
\midrule
Intercept              & 0.573 & 0.000 & 0.560 & 0.585 \\
Condition: Intervened  & 0.003 & 0.198 & -0.002 & 0.008 \\
Group Variance         & 0.044 &       &        &       \\
\bottomrule
\end{tabularx}}

\end{table}

\begin{table}[t]
\caption{LMEM predicting inter-person Beat Consistency for the audio delay manipulation. Delay set at \(0.75\)s.}
\label{tab:lmem-cross-bc-audio-delayx0.75}
\small
\centering
{\tabcolsep=2pt\def\arraystretch{1.0}
\begin{tabularx}{\linewidth}{@{}>{\arraybackslash}l *5{Y}}
\toprule
\textbf{Predictor} & \textbf{Coef.} & \textbf{p-value} & \multicolumn{2}{c}{\(\mathbf{95\%}\) \textbf{CI}} \\
\midrule
Intercept              & 0.571 & 0.000 & 0.559 & 0.583 \\
Condition: Intervened  & -0.003 & 0.422 & -0.009 & 0.004 \\
Group Variance         & 0.039 &       &       &       \\
\bottomrule
\end{tabularx}}

\end{table}

\begin{table}[t]
\caption{LMEM predicting inter-person Beat Consistency for the audio delay manipulation. Delay set at \(1.4\)s.}
\label{tab:lmem-cross-bc-audio-delayx1.4}
\small
\centering
{\tabcolsep=2pt\def\arraystretch{1.0}
\begin{tabularx}{\linewidth}{@{}>{\arraybackslash}l *5{Y}}
\toprule
\textbf{Predictor} & \textbf{Coef.} & \textbf{p-value} & \multicolumn{2}{c}{\(\mathbf{95\%}\) \textbf{CI}} \\
\midrule
Intercept                 & 0.565 & 0.000 & 0.553 & 0.577 \\
Condition: Intervened     & -0.008 & 0.090 & -0.017 & 0.001 \\
Group Variance            & 0.029 &       &        &       \\
\bottomrule
\end{tabularx}}

\vspace{-0.3cm}
\end{table}

\clearpage


\begin{table}[htbp]
\caption{LMEM predicting inter-person Soft-DTW distances. Dampening strength at \(10\).}
\label{tab:sdtw-inter-person-intervenedx10}
\small
\centering
{\tabcolsep=2pt\def\arraystretch{1.0}
\begin{tabularx}{\linewidth}{@{}>{\arraybackslash}l *5{Y}}
\toprule
\textbf{Predictor} & \textbf{Coef.} & \textbf{p-value} & \multicolumn{2}{c}{\(\mathbf{95\%}\) \textbf{CI}} \\
\midrule
Intercept                                           &  0.323 &     0.000  &  0.272 &  0.373  \\
Condition: Intervened                               & -0.637 &     0.000  & -0.688 & -0.586  \\
Joint (vs Right Hand): Left Arm                     & -0.059 &     0.023  & -0.110 & -0.008  \\
Joint (vs Right Hand): Left Hand                    & -0.003 &     0.906  & -0.054 &  0.048  \\
Joint (vs Right Hand): Right Arm                    & -0.047 &     0.073  & -0.097 &  0.004  \\
Intervened × Left Arm                               &  0.012 &     0.741  & -0.060 &  0.084  \\
Intervened × Left Hand                              &  0.028 &     0.450  & -0.044 &  0.100  \\
Intervened × Right Arm                              & -0.002 &     0.946  & -0.074 &  0.069  \\
Group Variance                                      &  0.376 &            &        &         \\
\bottomrule
\end{tabularx}}
\end{table}

\begin{table}[htbp]
\caption{LMEM predicting inter-person Soft-DTW distances. Dampening strength at $20$.}
\label{tab:sdtw-inter-person-intervenedx20}
\small
\centering
{\tabcolsep=2pt\def\arraystretch{1.0}
\begin{tabularx}{\linewidth}{@{}>{\arraybackslash}l *5{Y}}
\toprule
\textbf{Predictor} & \textbf{Coef.} & \textbf{p-value} & \multicolumn{2}{c}{\(\mathbf{95\%}\) \textbf{CI}} \\
\midrule
Intercept                                           &  0.320 &     0.000  &  0.269 &  0.371  \\
Condition: Intervened                               & -0.647 &     0.000  & -0.698 & -0.595  \\
Joint (vs Right Hand): Left Arm                     & -0.059 &     0.026  & -0.110 & -0.007  \\
Joint (vs Right Hand): Left Hand                    & -0.003 &     0.908  & -0.055 &  0.049  \\
Joint (vs Right Hand): Right Arm                    & -0.046 &     0.078  & -0.098 &  0.005  \\
Intervened × Left Arm                               &  0.026 &     0.485  & -0.047 &  0.099  \\
Intervened × Left Hand                              &  0.054 &     0.145  & -0.019 &  0.127  \\
Intervened × Right Arm                              &  0.011 &     0.758  & -0.061 &  0.084  \\
Group Variance                                      &  0.368 &            &        &         \\
\bottomrule
\end{tabularx}}

\end{table}

\begin{table}[htbp]
\caption{LMEM predicting inter-person Soft-DTW distances. Dampening strength at $30$.}
\label{tab:sdtw-inter-person-intervenedx30}
\small
\centering
{\tabcolsep=2pt\def\arraystretch{1.0}
\begin{tabularx}{\linewidth}{@{}>{\arraybackslash}l *5{Y}}
\toprule
\textbf{Predictor} & \textbf{Coef.} & \textbf{p-value} & \multicolumn{2}{c}{\(\mathbf{95\%}\) \textbf{CI}} \\
\midrule
Intercept                                           &  0.314 &     0.000  &  0.263 &  0.365  \\
Condition: Intervened                               & -0.631 &     0.000  & -0.683 & -0.579  \\
Joint (vs Right Hand): Left Arm                     & -0.058 &     0.028  & -0.110 & -0.006  \\
Joint (vs Right Hand): Left Hand                    & -0.003 &     0.910  & -0.055 &  0.049  \\
Joint (vs Right Hand): Right Arm                    & -0.046 &     0.083  & -0.098 &  0.006  \\
Intervened × Left Arm                               &  0.016 &     0.673  & -0.058 &  0.089  \\
Intervened × Left Hand                              &  0.067 &     0.077  & -0.007 &  0.140  \\
Intervened × Right Arm                              &  0.001 &     0.969  & -0.072 &  0.075  \\
Group Variance                                      &  0.360 &            &        &         \\
\bottomrule
\end{tabularx}}

\end{table}

\begin{table}[htbp]
\caption{LMEM predicting inter-person Soft-DTW distances. Dampening strength at $40$.}
\label{tab:sdtw-inter-person-intervenedx40}
\small
\centering
{\tabcolsep=2pt\def\arraystretch{1.0}
\begin{tabularx}{\linewidth}{@{}>{\arraybackslash}l *5{Y}}
\toprule
\textbf{Predictor} & \textbf{Coef.} & \textbf{p-value} & \multicolumn{2}{c}{\(\mathbf{95\%}\) \textbf{CI}} \\
\midrule
Intercept                                           &  0.307 &     0.000  &  0.257 &  0.358  \\
Condition: Intervened                               & -0.605 &     0.000  & -0.657 & -0.552  \\
Joint (vs Right Hand): Left Arm                     & -0.058 &     0.031  & -0.111 & -0.005  \\
Joint (vs Right Hand): Left Hand                    & -0.003 &     0.911  & -0.056 &  0.050  \\
Joint (vs Right Hand): Right Arm                    & -0.046 &     0.087  & -0.098 &  0.007  \\
Intervened × Left Arm                               & -0.006 &     0.875  & -0.080 &  0.068  \\
Intervened × Left Hand                              &  0.063 &     0.098  & -0.012 &  0.137  \\
Intervened × Right Arm                              & -0.020 &     0.594  & -0.095 &  0.054  \\
Group Variance                                      &  0.355 &            &        &         \\
\bottomrule
\end{tabularx}}

\end{table}

\begin{table}[htbp]
\caption{LMEM predicting inter-person Soft-DTW distances. Dampening strength at $50$.}
\label{tab:sdtw-inter-person-intervenedx50}
\small
\centering
{\tabcolsep=2pt\def\arraystretch{1.0}
\begin{tabularx}{\linewidth}{@{}>{\arraybackslash}l *5{Y}}
\toprule
\textbf{Predictor} & \textbf{Coef.} & \textbf{p-value} & \multicolumn{2}{c}{\(\mathbf{95\%}\) \textbf{CI}} \\
\midrule
Intercept                                           &  0.303 &     0.000  &  0.252 &  0.353  \\
Condition: Intervened                               & -0.575 &     0.000  & -0.627 & -0.522  \\
Joint (vs Right Hand): Left Arm                     & -0.054 &     0.042  & -0.106 & -0.002  \\
Joint (vs Right Hand): Left Hand                    & -0.004 &     0.894  & -0.056 &  0.049  \\
Joint (vs Right Hand): Right Arm                    & -0.042 &     0.112  & -0.094 &  0.010  \\
Intervened × Left Arm                               & -0.025 &     0.506  & -0.099 &  0.049  \\
Intervened × Left Hand                              &  0.058 &     0.124  & -0.016 &  0.132  \\
Intervened × Right Arm                              & -0.039 &     0.299  & -0.113 &  0.035  \\
Group Variance                                      &  0.353 &            &        &         \\
\bottomrule
\end{tabularx}}

\end{table}

\begin{table}[H]
\caption{Results of statistical comparison across varying pitch range limits.}
\label{tab:pitch-limits-wilcx}
\small
\centering
{\tabcolsep=20pt\def\arraystretch{1.0}
\begin{tabularx}{\linewidth}{@{}>{\arraybackslash}X r r@{}}
\toprule
\textbf{Limit/$\pm$ Hz} & \textbf{W-value} & \textbf{P-value} \\
\midrule
140 & 5974 & <0.05 \\
120 & 5934 & <0.05 \\
100 & 5630 & <0.05 \\
 80 & 4161 & <0.05 \\
 40 & 3972 & <0.05 \\
\bottomrule
\end{tabularx}}

\end{table}

\end{document}